\documentclass[epj]{svjour}

\usepackage{graphicx}
\usepackage{amsmath}
\usepackage{xspace}
\usepackage{float}
\usepackage{latexsym}

\usepackage{hyperref}
\usepackage{color}

\newcommand{\ord}{{\mathcal O}}

\newcommand{\be}{\begin{equation}}
\newcommand{\ee}{\end{equation}}
\newcommand{\bea}{\begin{eqnarray}}
\newcommand{\eea}{\end{eqnarray}}
\newcommand{\ba}{\begin{array}}
\newcommand{\ea}{\end{array}}

\def\siml{{\ \lower-1.2pt\vbox{\hbox{\rlap{$<$}\lower6pt\vbox{\hbox{$\sim$}}}}\ }}

\def\als{\alpha_{s}}
\def\al{\alpha}
\def\lQ{\Lambda_{\rm QCD}}

\newcommand{\nn}{\nonumber}

\newcommand{\eq}[1]{Eq.~\eqref{#1}}
\newcommand{\eqs}[2]{Eqs.~\eqref{#1} and \eqref{#2}}

\newcommand{\eqss}[2]{Eqs.~\eqref{#1}-\eqref{#2}}

\newcommand{\fig}[1]{Figure~\ref{#1}}
\newcommand{\figs}[2]{Figures~\ref{#1} and~\ref{#2}}
\newcommand{\figss}[3]{Figures~\ref{#1}, \ref{#2} and~\ref{#3}}
\newcommand{\app}[1]{Appendix~\ref{#1}}

\newcommand{\Sec}[1]{Section~\ref{#1}}

\newcommand{\rcite}[1]{Ref.~\cite{#1}}
\newcommand{\rcites}[1]{Refs.~\cite{#1}}

\def\dsl{\,\raise.15ex\hbox{/}\mkern-13.5mu D}

\newcommand{\MS}{\overline{\rm MS}}

\newcommand{\vp}{{\bf p}}

\newcommand{\bp}{{\mathbf p}}

\newcommand{\bk}{{\mathbf k}}
\newcommand{\br}{{\mathbf r}}
\newcommand{\bq}{{\mathbf q}}
\newcommand{\bE}{{\mathbf E}}
\newcommand{\bA}{{\mathbf A}}
\newcommand{\bL}{{\mathbf L}}

\newcommand{\eps}{\epsilon}

\def\lQ{\Lambda_{\rm QCD}}
\def\al{\alpha}
\def\als{\alpha_{\rm s}}
\def\siml{{\ \lower-1.2pt\vbox{\hbox{\rlap{$<$}\lower6pt\vbox{\hbox{$\sim$}}}}\ }} 

\def\lla{\langle\!\langle}
\def\rra{\rangle\!\rangle}
\newcommand{\Appendix}[1]%
    {%
     \section{#1}%
      }
    
\newcommand{\beq}{\begin{equation}}
\newcommand{\eeq}{\end{equation}}

\newcommand{\df}{\mathrm{d}}

\allowdisplaybreaks[3]

\title{Relativistic corrections to the static energy in terms of Wilson loops at 
weak coupling}
\author{Clara Peset\inst{1}, Antonio Pineda\inst{1}, \and Maximilian Stahlhofen\inst{2}
}                     
\institute{Grup de F\'\i sica Te\`orica, Dept. F\'\i sica and IFAE-BIST, 
Universitat Aut\`onoma de Barcelona, \\ E-08193 Bellaterra (Barcelona), Spain 
\and PRISMA Cluster of Excellence, Institute of Physics, Johannes 
Gutenberg University, 55128 Mainz, Germany}
\abstract{
We consider the $\ord(1/m)$ and the spin-independent momentum-dependent 
$\ord(1/m^2)$ quasi-static energies of heavy quarkonium (with unequal masses). 
They are defined nonperturbatively in terms of Wilson loops.
We determine their short-distance behavior through ${\cal O}(\al^3)$ 
and ${\cal O}(\al^2)$, respectively.
In particular, we calculate the ultrasoft contributions to the quasi-static energies, which requires the resummation of potential interactions.
Our results can be directly compared to lattice simulations. 
In addition, we also compare the available lattice data with the expectations 
from effective string models for the long-distance behavior of the quasi-static 
energies.	
\keywords {NRQCD, potentials, heavy quarkonium}
\PACS{12.38.Bx, 12.39.Hg, 12.38.Cy, 14.40.Pq
  } 
}

\begin{document}

\authorrunning{Peset, Pineda, Stahlhofen}
\titlerunning{Relativistic corrections to the static energy in terms of Wilson 
loops at 
weak coupling}
\maketitle




\section{Introduction}

The near-threshold dynamics of quark-antiquark systems with large masses can be described by a properly generalized nonrelativistic (NR) Schr\"odinger equation. 
Such equation can be derived from first principles using effective field 
theories (EFT's) like potential NRQCD 
(pNRQCD)~\cite{Pineda:1997bj,Brambilla:1999xf} (for reviews 
see~\cite{Brambilla:2004jw,Pineda:2011dg}). This EFT systematically exploits the 
nonrelativistic scale hierarchy $m \gg mv \gg mv^2 \ldots\,$, where $v$ is the 
heavy-quark velocity in the center of mass frame. In the case of unequal 
heavy quark masses, we will assume $m_1\sim m_2 \sim m$.

In this paper we will mostly work in the weak-coupling regime, where the EFT can be schematically summarized as
\begin{align}
\,\left.
\begin{array}{ll}
&
\left(i\partial_0-\frac{{\bf p}^2}{  2 m_r}-V_s(r)\right)\phi(\br)=0
\\
&
+\; \text{interactions with other low-energy}
\\ &\hspace*{0.28cm} \text{ degrees of freedom}
\end{array} \right\} 
\text{pNRQCD}\,.
\label{pNRQCD}
\end{align}
Here, $m_r=m_1m_2/(m_1+m_2)$, $\phi(\br)$ represents the quarkonium wave 
function, $\bp$ is the momentum operator conjugate to the distance vector $\br$ 
between the  heavy quarks and $V_s$ is the color-singlet potential. We denote the color-octet potential by $V_o$. 
The potential $V_s$ consists of the static potential 
$V^{(0)}$ and its relativistic corrections, which are suppressed by 
inverse powers of the heavy quark masses. 
For the general case of unequal masses we have
\begin{align}
V_s =& 
V^{(0)} + \frac{V^{(1,0)} }{ m_1}+\frac{V^{(0,1)}}{ m_2}\nn\\
+& \frac{V^{(2,0)}}{ m_1^2}+ \frac{V^{(0,2)}}{ m_2^2}+\frac{V^{(1,1)}}{ m_1m_2}+\cdots,
\label{Vsexp}
\end{align}
where the spin-independent (SI) $1/m^2$ potentials are
\begin{align}
V^{(2,0)}_{SI} =& \frac{1}{ 2}\left\{{\bf p}_1^2,V_{{\bf p}^2}^{(2,0)}(r)\right\}+ V_r^{(2,0)}(r)
+V_{{\bf L}^2}^{(2,0)}(r)\frac{{\bf L}_1^2 }{ r^2} \,,
\label{V20SI}
\\
V^{(0,2)}_{SI} =& \frac{1 }{ 2}\left\{{\bf p}_2^2,V_{{\bf p}^2}^{(0,2)}(r)\right\}+ V_r^{(0,2)}(r)
+V_{{\bf L}^2}^{(0,2)}(r)\frac{{\bf L}_2^2 }{ r^2}\,,
\\
V^{(1,1)}_{SI} =& -\frac{1 }{ 2}\left\{{\bf p}_1\cdot {\bf p}_2,V_{{\bf p}^2}^{(1,1)}(r)\right\}+ V_r^{(1,1)}(r)\nn\\
&-V_{{\bf L}^2}^{(1,1)}(r)\frac{({\bf L}_1\cdot{\bf L}_2+ {\bf L}_2\cdot{\bf L}_1) }{ 2r^2}
\,,
\label{V11SI}
\end{align}
with ${\bf L}_1 \equiv {\bf r} \times {\bf p}_1$, ${\bf L}_2 \equiv {\bf r} 
\times {\bf p}_2$, and ${\bf p}_1$, ${\bf p}_2$ being the momenta of the heavy 
quarks in the center of mass frame. In what follows we will not consider spin-dependent potentials.

The potential in \eq{pNRQCD} is obtained by matching pNRQCD to NRQCD~\cite{Caswell:1985ui,Bodwin:1994jh}. 
There are several ways to perform this matching in practice. 
In general, the resulting expressions for the relativistic corrections to the 
potential at different orders in the $1/m$ expansion, also referred to 
as (higher order) `potentials', depend on the matching scheme. Nevertheless 
physical observables are of course unaffected. For a detailed discussion see \rcite{Peset:2015vvi}.

In this paper we focus on the potentials obtained from Wilson-loop 
matching.%
\footnote{ In the on-shell matching scheme for the equal mass case they 
can be found in \rcite{Kniehl:2002br} with N$^3$LO precision in the pNRQCD 
power counting, except for the static potential, which is given in 
\rcites{Anzai:2009tm,Smirnov:2009fh}. Previous partial results are given in 
\rcites{Gupta:1982qc,Kniehl:2001ju}.
The expressions in the unequal mass case have been derived in 
\rcite{Peset:2015vvi}.}
This scheme requires certain gauge invariant NRQCD and pNRQCD Green functions in position space to be equal at a given order in $1/m$.
On the NRQCD side these Green functions correspond to rectangular Wilson loops with chromo-electric/-magnetic field operator insertions.
The potentials are Wilson coefficients of pNRQCD operators.
Appropriate correlation functions of these operators are then matched onto the Wilson loops to determine the potentials.
For details on the  Wilson-loop matching we refer to \rcites{Brambilla:2000gk,Pineda:2000sz}. In the static limit, i.e.~at lowest order in $1/m$, the potential reduces to the original  Wilson-loop expression for the static potential $V^{(0)}$~\cite{Wilson:1974sk}. The Wilson-loop expression for the $1/m$ potential was obtained in \rcite{Brambilla:2000gk}, for the $1/m^2$ potential without light fermions in  \rcite{Pineda:2000sz}, and including light fermions in \rcite{Peset:2015vvi}.
The corresponding expressions in terms of Wilson loops for the SI momentum-dependent $1/m^2$ potentials, $V_ {{\bp}^2}$ and $V_ {{\bL}^2}$, were first derived in \rcite{Barchielli:1986zs}.                                                                                              The latter reference however missed the $1/m$ potential%
\footnote{as well as the SI momentum-independent $1/m^2$  potential $V_r$, which we do not consider here.}, which we also study in this paper.

Only (off-shell) soft modes, i.e. those modes with energies and momenta of 
order $1/r$, contribute to the potential in the Wilson loop scheme.
This is achieved by Taylor expanding in powers 
of $1/m$ before integrating over the gauge and light-quark degrees of freedom. 
 Thus, soft modes are properly accounted for, but ultrasoft modes still need to be eliminated.
At weak-coupling this is achieved by evaluating
the Wilson loops strictly in perturbation theory. Hence, the only
physical scale left in the Wilson-loop calculation of the potential is $1/r$.

Still, the definition of the potential in terms of Wilson loops naturally 
suggests its generalization beyond perturbation theory. We will refer to this 
generalized potential as the quasi-static energy.\footnote{In analogy to 
the potentials, we refer to the different (higher-order relativistic 
correction) terms in the $1/m$ expansion of the total quasi-static energy $E_s$, 
c.f. \eq{Es}, as the 'quasi-static energies'.}
To distinguish it from the potential $V$ we will use the symbol $E$ for the quasi-static energy.
The definition in terms of Wilson loops has the same form for both quantities, 
but $E$ is evaluated nonperturbatively, for instance through lattice 
simulations. Varying the distance $r$ allows us to move continuously from the 
extreme weak-coupling limit ($r \rightarrow 0$) to the strong-coupling regime 
($r \rightarrow \infty$) of the quasi-static energies.  
They are therefore ideal observables to study the interplay between both regimes in a controlled manner.

At long distances, the quasi-static energies are suitable for the study of nonperturbative QCD dynamics, e.g. by comparing different models with lattice simulations. In this regime there is a direct connection to the strong-coupling version of pNRQCD, as (omitting possible ultrasoft modes of chiral origin and treating $V^{(0)}$ as small compared to the soft scale) the total quasi-static energy corresponds to the Hamiltonian operator that describes the heavy quarkonium bound state, see \rcites{Brambilla:2000gk,Pineda:2000sz}. 
The determination of the quasi-static energy might therefore, for example, enable predictions of the higher excitations of bottomonium and charmonium. For a recent attempt in this direction see \rcite{Brambilla:2014eaa}.

Even at short distances, the regime we will focus on in this paper, the behavior of the quasi-static energies cannot be obtained from a strictly perturbative computation in the strong coupling $\al \equiv \als$ alone. 
The reason is that the so-called ultrasoft scale $\Delta V \equiv V^{(0)}_o-V^{(0)} \simeq C_A \al/(2r) \ll 1/r$ is generated dynamically by the exchange of an arbitrary number of Coulomb gluons. Including the ultrasoft effects requires a proper resummation of such potential interactions.
Typically this is done with the help of EFTs, but it can also be done using a diagrammatic approach, as we will show. Thus, in addition to the perturbative soft part $V_s$, the quasi-static energy $E_s$ also includes ultrasoft contributions, which are inaccessible in perturbation theory, even at weak coupling. 
Therefore it differs from the potential, which enters the Schr\"odinger equation that describes heavy quarkonium dynamics. 
Still, both objects are obtained by expanding in powers of $1/m$ before the integration over the relevant gauge or light-quark dynamical modes.

In \rcite{Peset:2015vvi}, we determined the soft contributions to the quasi-static energies up to $\mathcal{O}(\alpha^3/m)$ and $\mathcal{O}(\alpha^2/m^2)$. For the complete results at short distances we also need the ultrasoft contributions. 
We will compute them in this paper for the $1/m$ and the SI momentum-dependent $1/m^2$ quasi-static energies ($E_ {{\bp}^2}$, $E_ {{\bL}^2}$). 
Adding soft and ultrasoft contributions we obtain the $1/m$ quasi-static energy with ${\cal O}(\al^3)$ precision and $E_ {{\bp}^2}$, $E_ {{\bL}^2}$ with ${\cal O}(\al^2)$ precision. We also discuss the possible resummation of large logarithms for these quantities.

Our results can be directly confronted with the most recent lattice data from \rcites{Koma:2006si,Koma:2007jq,Koma:2009ws}.%
\footnote{For older simulations see~\rcite{Bali:1997am}.} 
One motivation for such a comparison is to perform quantitative checks of ultrasoft effects. 
The static energy $E^{(0)}$ starts to be sensitive to the ultrasoft scale only at ${\cal O}(\al^4)$, i.e. N$^3$LO~\cite{Appelquist:1977es,Brambilla:1999qa}.%
\footnote{For lower space-time dimension this happens at lower orders in the loop expansion. 
For instance in $2+1$ dimensions ultrasoft effects occur already at NNLO~\cite{Pineda:2010mb}.}
The high $\al$ suppression makes it difficult to study ultrasoft physics with this observable.
For the quasi-static energies $E^{(1,0)}$, $E^{(0,1)}$ and $E^{(1,1)}$ ultrasoft effects already contribute at NLO, for  $E^{(2,0)}$ and $E^{(0,2)}$  even at leading non-vanishing order.
They are therefore a good testing ground for both the ultrasoft physics and the reliability of the weak coupling prediction at a given distance $r$.  
This is important for several reasons. 
For example to give support to perturbative analyses of the heavy quarkonium spectrum that include ultrasoft contributions at weak coupling~\cite{Penin:2002zv,Ayala:2014yxa,Kiyo:2015ufa}. 
It is also crucial for determinations of the strong coupling constant $\al$ from lattice simulations of the static energy~\cite{Bazavov:2012ka}, as they rely on the correct treatment of ultrasoft effects over the whole range of (short) distances they probe. Finally, we would also like to remark that in ${\cal N}=4$ supersymmetric Yang-Mills theories the static energy $E^{(0)}$ is also sensitive to ultrasoft effects starting at NLO, see \cite{Erickson:1999qv,Pineda:2007kz,Stahlhofen:2012zx,Prausa:2013qva}.
Preliminary results from lattice simulations for this quantity have been reported in~\cite{Schaich:2016jus}.
They qualitatively agree with the perturbative predictions (including the ultrasoft contributions) at small distances.
A more detailed analysis would however be welcome.

\section{The quasi-static energy and its relativistic corrections}
\label{sec:Wilson}

We will use the following definitions for the Wilson-loop operators and their vacuum expectation values. 
The angular brackets $\langle \dots \rangle$ stand for the average value over the
QCD action, $W_\Box$ for the rectangular static Wilson loop of
dimensions $r\times T_W$,
\begin{align}
W_\Box \equiv {\rm P} \exp\left\{{\displaystyle - i g \oint_{r\times T_W} \!\!dz^\mu A_{\mu}(z)}\right\},
\end{align}
and 
\begin{align}
\langle\!\langle \dots \rangle\!\rangle
\equiv \frac{ \langle  {\rm P} \dots W_\Box\rangle }{  \langle  W_\Box\rangle} \,,
\label{WLcorrelatorDef}
\end{align}
where P is the path-ordering operator.
Moreover, we define the {\it connected} Wilson loop average
\begin{align}
&\lla O_1(t_1)O_2(t_2)\rra_c= \lla O_1(t_1)O_2(t_2)\rra 
-\lla O_1(t_1)\rra\lla O_{2}(t_2)\rra 
\label{ConnectedWLdef}
\end{align} 
of operator insertions $O_1$, $O_2$ in the Wilson (static quark) lines at times $t_1$ and $t_2$ with $T_W/2 \ge t_1 \ge t_2 \ge  -T_W/2$. 
We also define the short-hand notation 
\begin{align}
\lim_{T\rightarrow \infty} \equiv \lim_{T\rightarrow \infty}\lim_{T_W\rightarrow \infty}
\,,
\end{align}
where $T_W$ is the time length of the Wilson loop and $T$ the upper limit of the
time integrals in the definitions below. 
By taking first the $T_W\rightarrow \infty$ limit, the averages $\lla \dots \rra$ 
become independent of $T_W$ and thus invariant under global time translations, 
see \rcite{Pineda:2000sz} for details. 

The total color-singlet quasi-static energy is defined as the sum of the kinetic terms and the 
nonperturbative generalization of the potential in terms of Wilson loops. It reads  
\begin{align}
\label{Es}
&E_s({\bf r}, {\bf p}, {\bf P}_{\bf R}, {\bf S}_1,{\bf S}_2) = \nn\\
&\qquad \frac {{\bf p}^2 }{ 2\, m_{\rm r}}
+ 
\frac{{\bf P}_{\bf R}^2 }{ 2\, M} + 
E^{(0)} + \frac{E^{(1,0)} }{ m_1}+\frac{E^{(0,1)}}{ m_2}\nn\\
&\qquad
+ \frac{E^{(2,0)} }{ m_1^2}+ \frac{E^{(0,2)}}{ m_2^2}+\frac{E^{(1,1)} }{ m_1m_2}+ \ord\bigg(\frac{1}{m^3} \bigg)\,,
\end{align}
where ${\bf P}_{\bf R}$ is the center of mass momentum operator, $\bp=-i \nabla_\br$, $M=m_1+m_2$, the reduced mass $m_r = m_1 m_2/M$, and 
\begin{align}
E^{(0)}(r)  =  \lim_{T\to\infty}\frac{i}{ T} \ln \langle W_\Box 
\rangle
\label{v0}
\end{align}
is the static energy~\cite{Wilson:1974sk}.
The formal definitions for the higher order quasi-static energies to $\ord(1/m)$ can be found in \rcite{Brambilla:2000gk} and to $\ord(1/m^2)$ in \rcites{Pineda:2000sz,Peset:2015vvi}.
Here we recall the ones relevant for the present work.
At $\ord(1/m)$ we have
\begin{align}
E^{(1,0)}(r)=
-\frac{1}{ 2} \lim_{T\rightarrow \infty}\int_0^{T}dt \, t \, \lla g{\bf E}_1(t)\cdot g{\bf E}_1(0) \rra_c\,.
\label{E1}
\end{align}
The SI momentum-dependent $1/m^2$ potentials are parametrized in complete analogy to \eqss{V20SI}{V11SI} and given by
\begin{align}
\label{Ep2}
E_{{\bf p}^2}^{(2,0)}(r&)=\frac{i}{ 2}{\hat {\bf r}}^i{\hat {\bf r}}^j
\lim_{T\rightarrow \infty}\int_0^{T}dt \,t^2 \lla g{\bf E}_1^i(t) g{\bf E}_1^j(0) \rra_c
\,,
\\
\label{EL2}
E_{{\bf L}^2}^{(2,0)}(r)&=\frac{i}{ 2(d-1)}
 \left(\delta^{ij}-d{\hat {\bf r}}^i{\hat {\bf r}}^j \right)\nn\\
&\times\lim_{T\rightarrow \infty}\int_0^{T}dt \, t^2 \lla g{\bf E}_1^i(t) g{\bf E}_1^j(0) \rra_c
\,,
\\
\label{Ep211}
E_{{\bf p}^2}^{(1,1)}(r)&=i\,{\hat {\bf r}}^i{\hat {\bf r}}^j
\lim_{T\rightarrow \infty}\int_0^{T}dt \, t^2
\lla g{\bf E}_1^i(t) g{\bf E}_2^j(0) \rra_c
\,,
\\
\label{EL211}
E_{{\bf L}^2}^{(1,1)}(r)&= \frac{i}{d-1}\left( \delta^{ij}-d{\hat {\bf r}}^i{\hat {\bf r}}^j \right)\nn\\
&\times\lim_{T\rightarrow \infty}\int_0^{T}dt \,t^2
\lla g{\bf E}_1^i(t) g{\bf E}_2^j(0) \rra_c 
\,,
\end{align}
where $d=D-1=3+2\epsilon$, ${\hat {\bf r}} = {\bf r}/r$ and $\bE_n$ is the 
chromo-electric field-strength operator to be inserted on the Wilson line 
associated with the static quark ($n=1$) or on the Wilson line associated with 
the static antiquark ($n=2$).
The coupling $g$ in the above expressions will be replaced by the bare coupling $g_B$ in the actual calculations carried out in the next section.
We do not consider spin (${\bf S}_{1,2}$)-dependent or momentum ($\bp_{1,2}$)-independent quasi static energies in this paper.

The potential $V_s$, which enters the Schr\"odinger equation that describes heavy quarkonium dynamics, is determined by expanding in powers of $1/m$ before integrating over any gauge and 
light-quark dynamical degrees of freedom. 
For the ultrasoft scales this implies $\Delta V \gg \bp^2/m$.
Concerning the ultrasoft contributions to the energy of a physical bound state of dynamical heavy quarks such a $1/m$ expansion would be incorrect, as according to the Virial theorem  $\Delta V \sim \bp^2/m$. 
This is the reason why ultrasoft modes are removed from the potential.
They are retained in the quasi-static energy, which can therefore be (formally) regarded as the heavy quarkonium energy in the quasi-static limit: $\Delta V \gg \bp^2/m$.

The relation between the total (singlet) quasi-static energy $E_s$ and the potential in the Wilson loop scheme $V_{s,W}$ is
\begin{align}
\label{EsplusEus}
E_{s}(r)=V_{s,W}(r)+\delta E_{s,{\rm US}}(r)
\,,
\end{align}
where $\delta E_{s,{\rm US}}$ denotes the ultrasoft contribution to the  quasi-static energy.
The analogous splitting into the corresponding (soft) potential and ultrasoft contribution is understood for each of the quasi-static energies $E^{(\ldots)}_{\ldots}$ in \eqss{v0}{EL211}. 

In order to compute the potential $V_{s,W}(r)$ in perturbation theory, the ultrasoft contribution must be completely subtracted. The latter can be achieved in dimensional regularization by expanding in the ultrasoft scale $\Delta V$ {\it before} performing the loop integrations.
As $\Delta V \propto \al$ this means we simply have to ensure a strictly perturbative calculation in $\al$.
Thus, only the soft scale $1/r$ is left and the loop integrals become homogeneous in that scale.
The potentials then take the form of a power series in $\al \propto g^2$ (and, eventually, in $\Delta V$, when working beyond the order we work at in this paper). 
In summary we have,
\begin{align}
	V_{s,W}(r)=E_s(r)|_{\rm soft}\,.
	\label{VsW}
\end{align}
We have presented the results for the potential $V_{s,W}$ up to $\ord(\al^3/m)$ and $\ord(\al^2/m^2)$ in \rcite{Peset:2015vvi}.

In the weak-coupling regime ultrasoft degrees of freedom,
with energies and momenta of order $\Delta V \equiv V^{(0)}_o-V^{(0)} \sim 
C_A\al/r$, 
contribute to the quasi-static energies through $\delta E_{s,{\rm US}}$. These contributions to the Wilson loops correspond to the contributions from the lower (ultrasoft) energy regions in an asymptotic expansion of the loop integrations according to the method of regions~\cite{Beneke:1997zp}. This implies that (some of) the loop integrals are sensitive to the ultrasoft scale.
In analogy to \eq{VsW} we write
\begin{align}
\delta E_{s,{\rm US}}(r)=E_s(r)|_{\rm US}\,.
\end{align}
Next, we explain how to compute $\delta E_{s,{\rm US}}$ in detail.

\subsection{Ultrasoft calculation}
\label{sec:E}

\begin{figure}[t]
	\includegraphics[width=0.25\columnwidth]{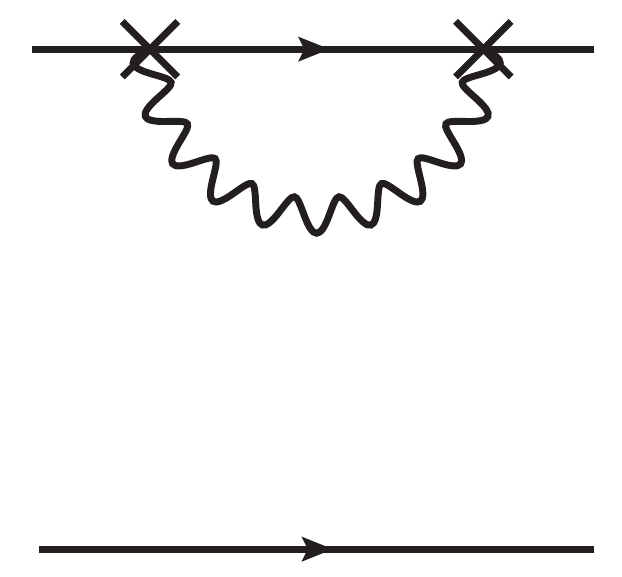}
	\put(-45,-10){(1a)} 
	\includegraphics[width=0.25\columnwidth]{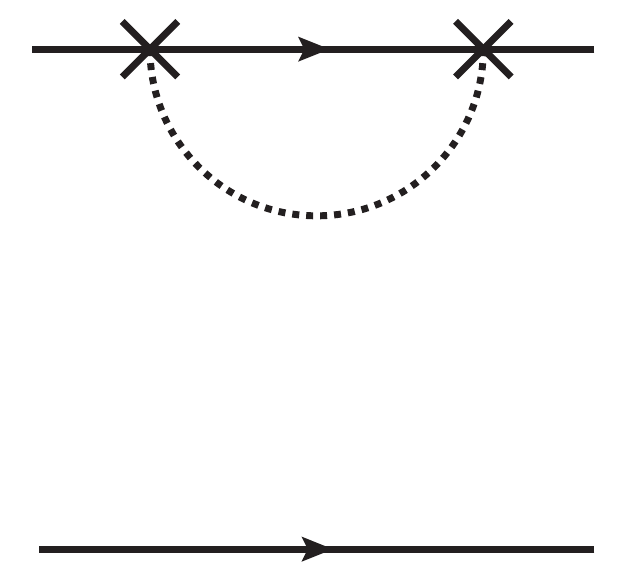}
	\put(-45,-10){(1b)} 
	\includegraphics[width=0.25\columnwidth]{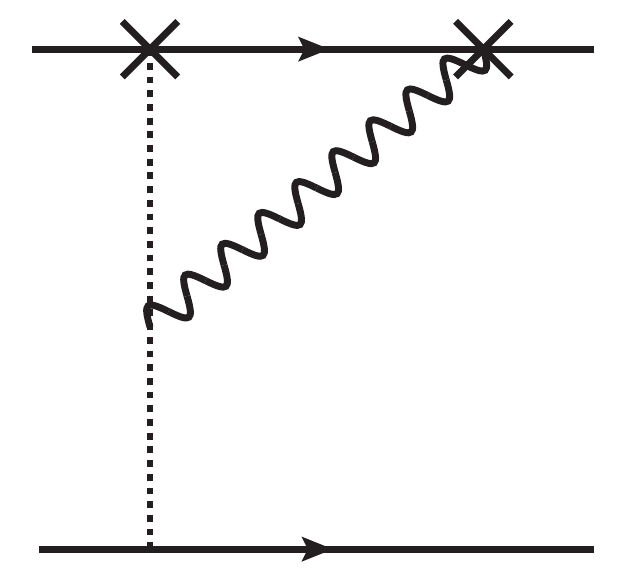}
	\put(-45,-10){(2a)} 
	\includegraphics[width=0.25\columnwidth]{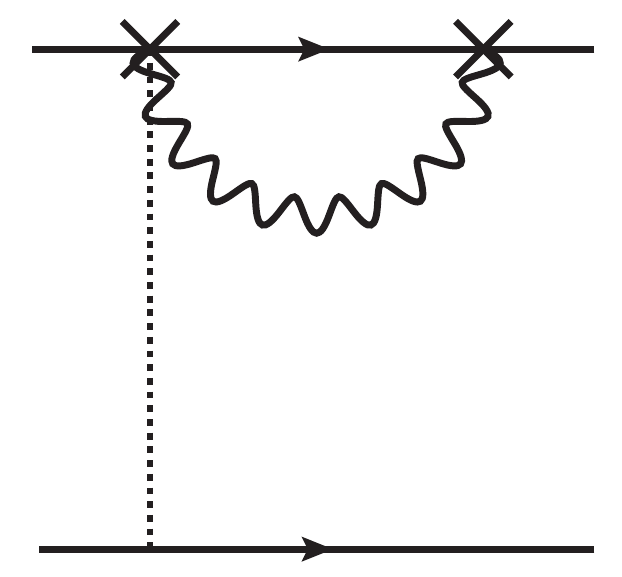}
	\put(-45,-10){(2b)} 
	\vspace{1ex}
	\includegraphics[width=0.25\columnwidth]{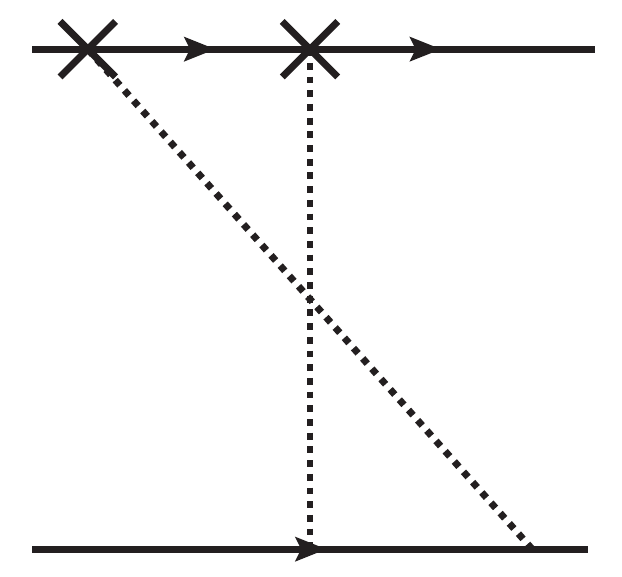}
	\put(-45,-10){(2c)} 
	\includegraphics[width=0.25\columnwidth]{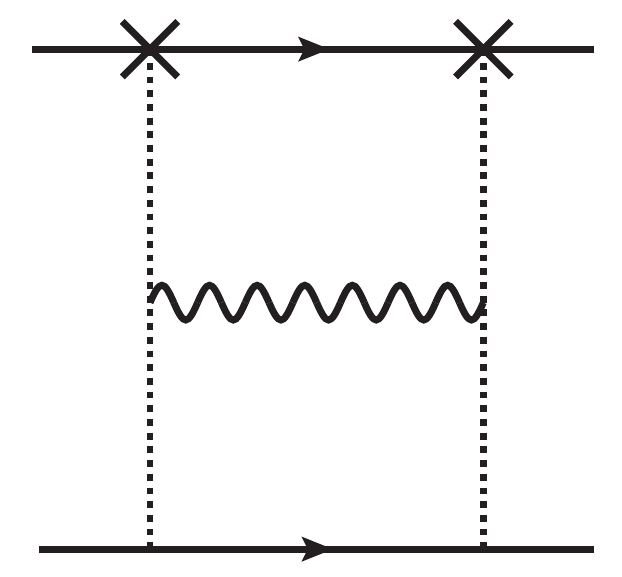} 
	\put(-45,-10){(3a)}
	\includegraphics[width=0.25\columnwidth]{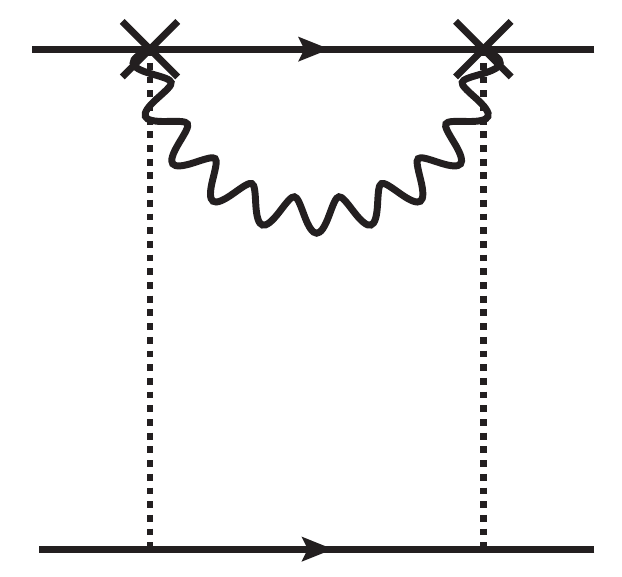} 
	\put(-45,-10){(3b)}
	\includegraphics[width=0.25\columnwidth]{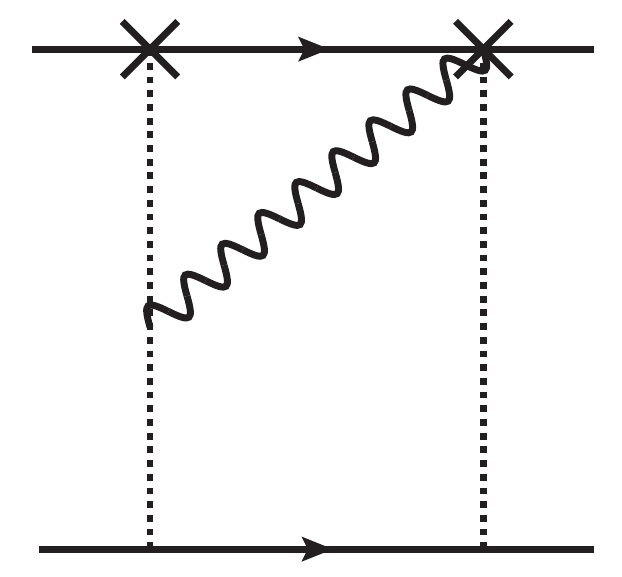}
	\put(-45,-10){(3c)}
	\vspace{1ex}
	\includegraphics[width=0.25\columnwidth]{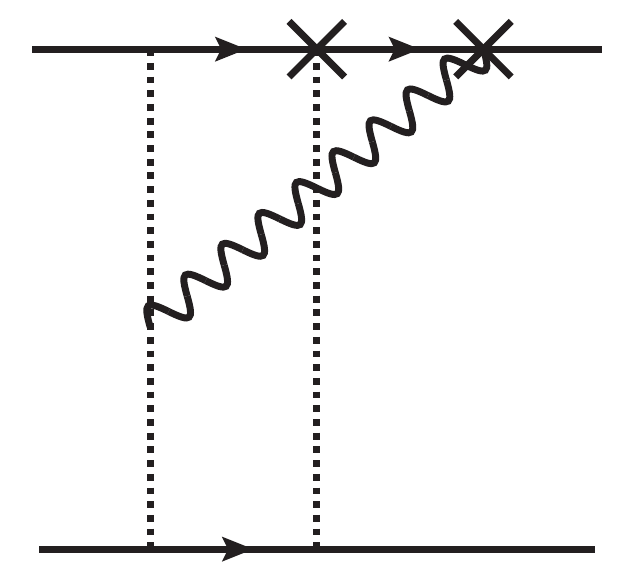}
	\put(-45,-10){(3d)}
	\includegraphics[width=0.25\columnwidth]{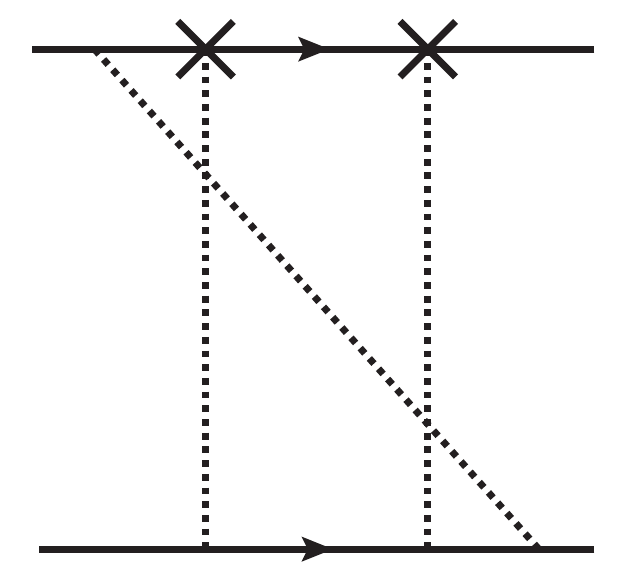}
	\put(-45,-10){(3e)}
	\includegraphics[width=0.25\columnwidth]{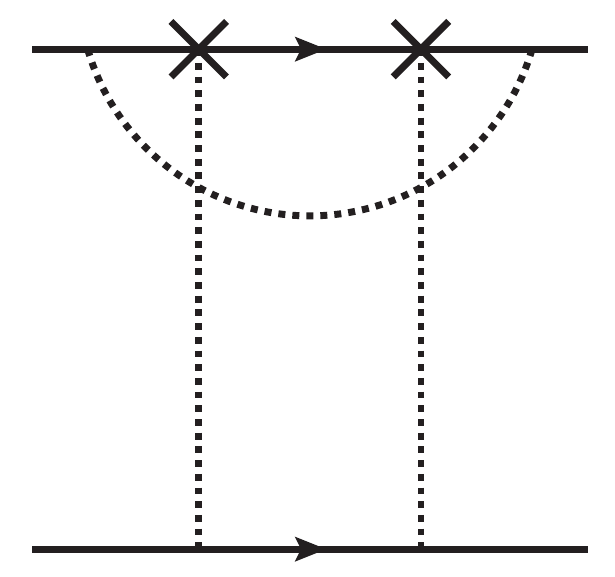}
	\put(-45,-10){(3f)}
	\includegraphics[width=0.25\columnwidth]{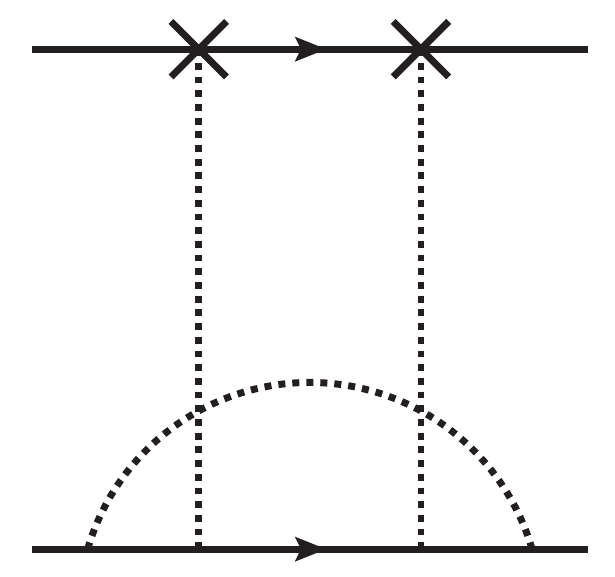}
	\put(-45,-10){(3g)}%
	\vspace{1ex}
	\includegraphics[width=0.25\columnwidth]{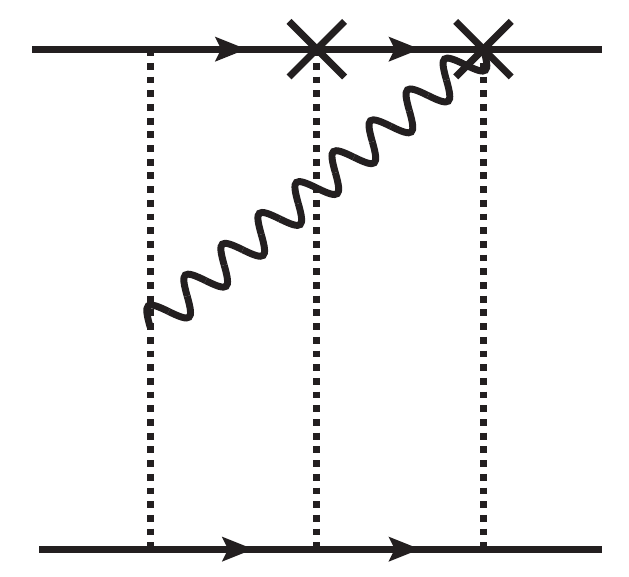}
	\put(-45,-10){(4a)}
	\includegraphics[width=0.25\columnwidth]{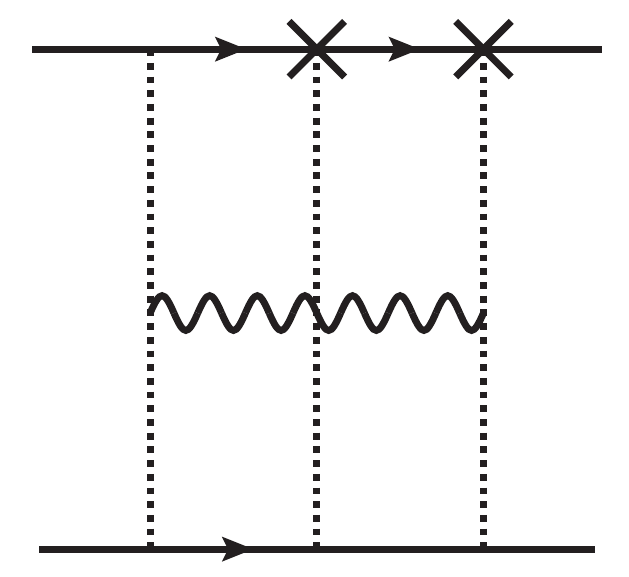} 
	\put(-45,-10){(4b)}
	\includegraphics[width=0.25\columnwidth]{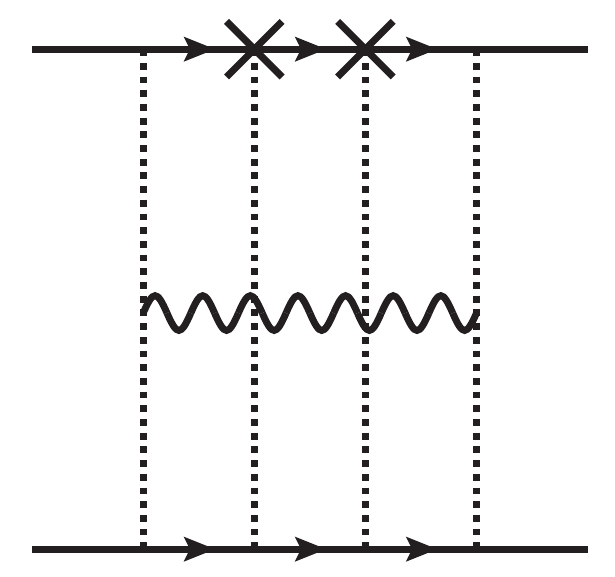} 
	\put(-45,-10){(5a)}
\caption{Connected (2-Wilson-line irreducible) diagrams with one ultrasoft 
gluon exchange contributing to $E_{\rm US}^{(2,0)}$ and $E_{\rm US}^{(1,0)}$.
Left-right mirror graphs are understood. 
Dotted lines represent the $A^0$, wiggled lines the $\bA$ components of a 
gluon. The time axis in these diagrams extends from left to right.
The crosses symbolize the chromo-electric operator insertions at times 
$0$ and $t$ on the upper Wilson line (solid).
Vertical lines are associated with potential gluons. The ultrasoft gluon is 
depicted as diagonal, horizontal or curvy line.
We adopt the NRQCD convention for the fermion flow (little arrows on the 
Wilson lines) explained in \fig{EFeynRules}.
Diagrams $(1b)$, $(2c)$, $(3e)$, $(3f)$, $(3g)$ vanish in Coulomb gauge as they are proportional to scaleless integrals.
\label{Vx0Diags}}   
\end{figure}

\begin{figure}[t]
	\includegraphics[width=0.25\columnwidth]{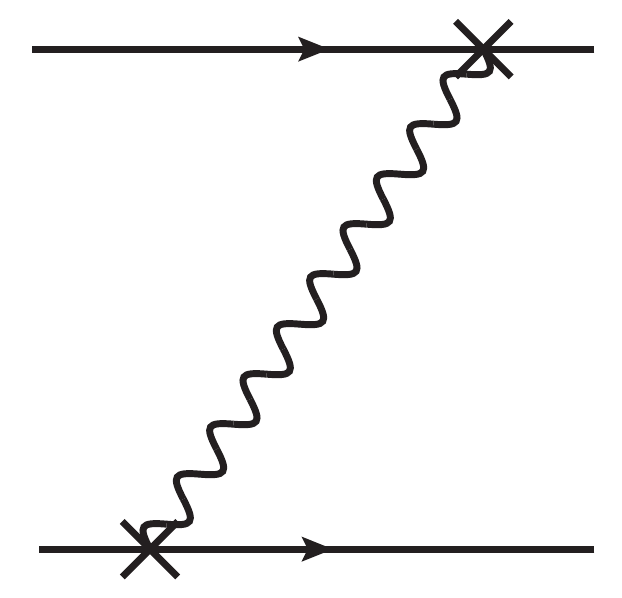}
	\put(-45,-10){(1a)} 
	\includegraphics[width=0.25\columnwidth]{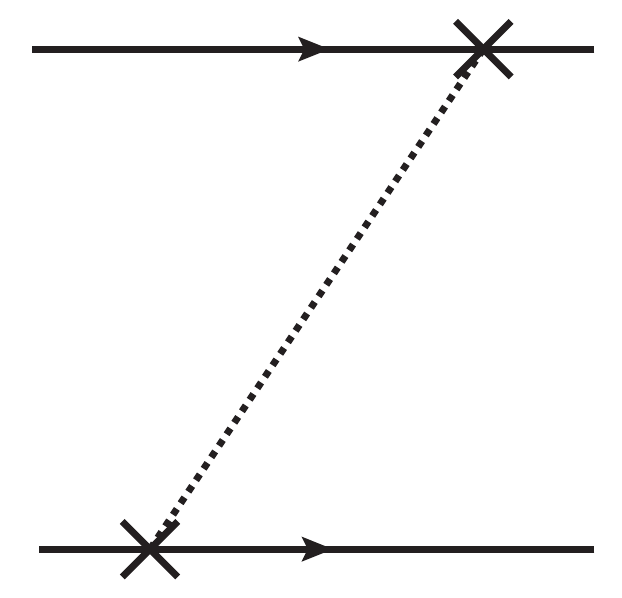}
	\put(-45,-10){(1b)} 
	\includegraphics[width=0.25\columnwidth]{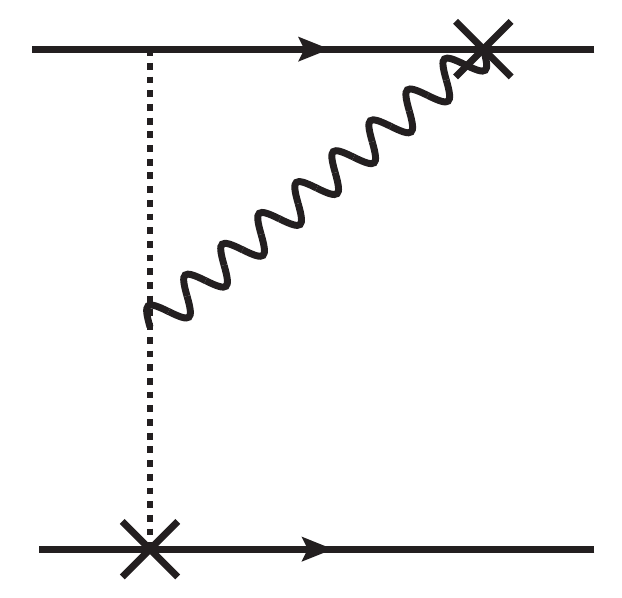}
	\put(-45,-10){(2a)} 
	\includegraphics[width=0.25\columnwidth]{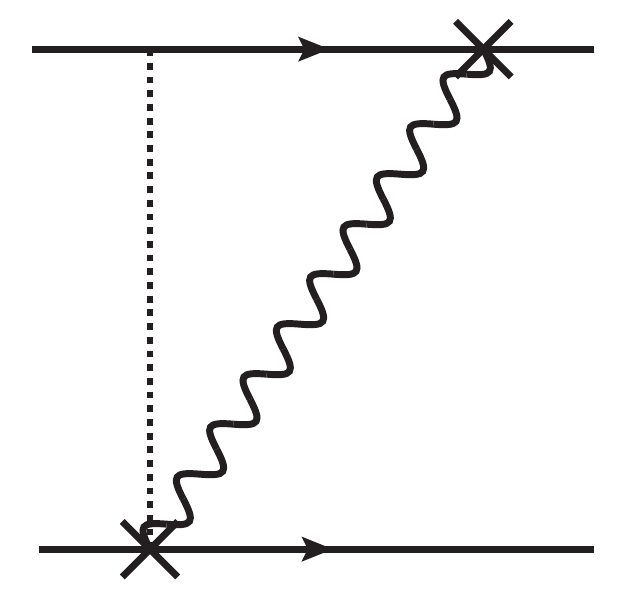}
	\put(-45,-10){(2b)} 
	\vspace{1ex}
	\includegraphics[width=0.25\columnwidth]{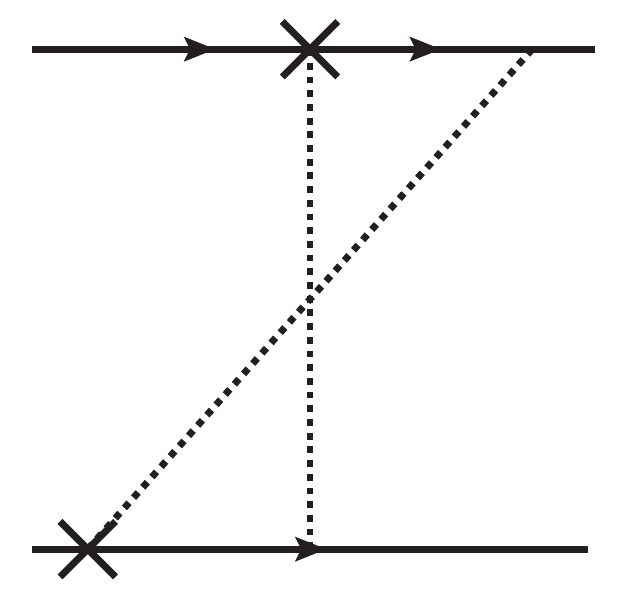}
	\put(-45,-10){(2c)} 
	\includegraphics[width=0.25\columnwidth]{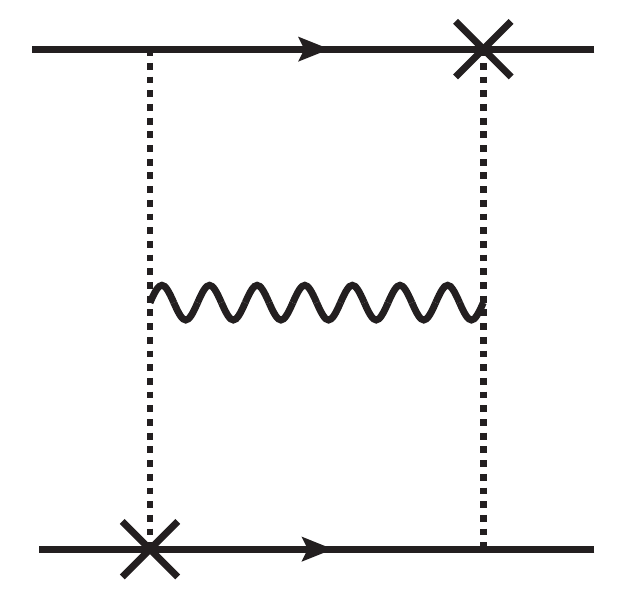} 
	\put(-45,-10){(3a)}
	\includegraphics[width=0.25\columnwidth]{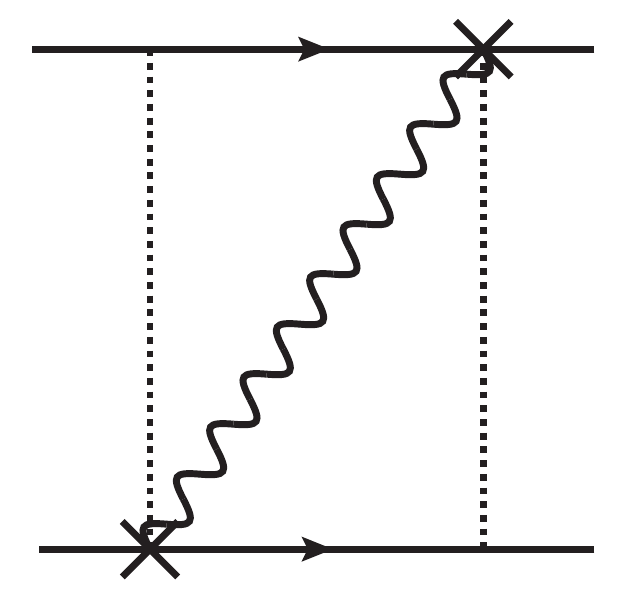} 
	\put(-45,-10){(3b)}
	\includegraphics[width=0.25\columnwidth]{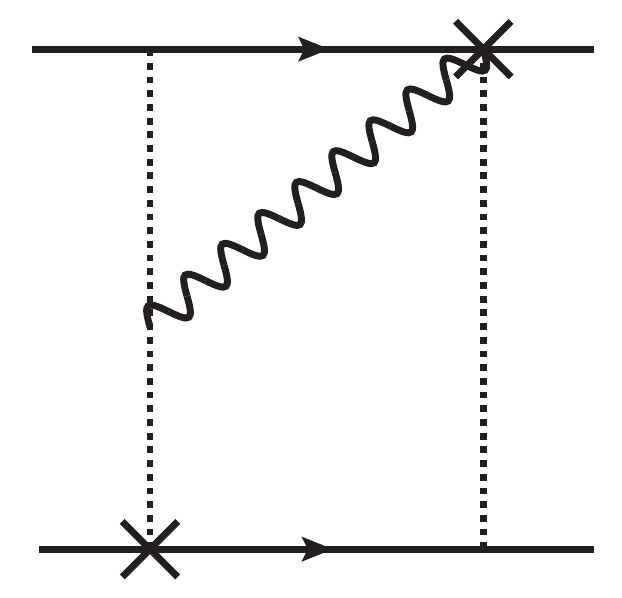}
	\put(-45,-10){(3c)}
	\vspace{1ex}
	\includegraphics[width=0.25\columnwidth]{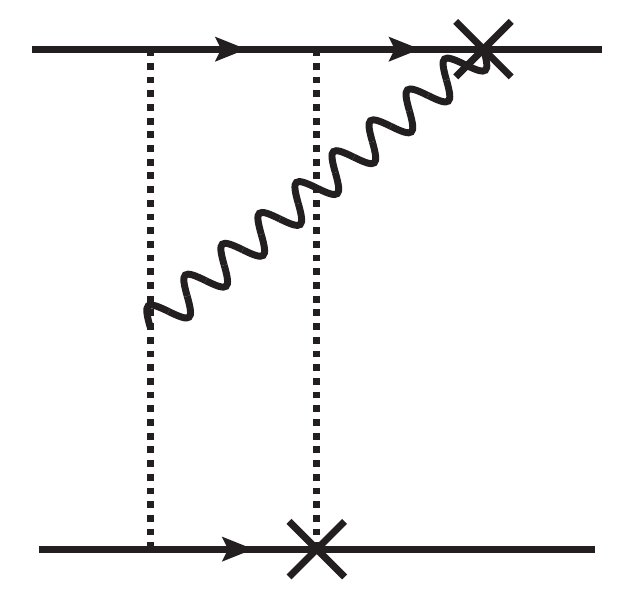}
	\put(-45,-10){(3d)}
	\includegraphics[width=0.25\columnwidth]{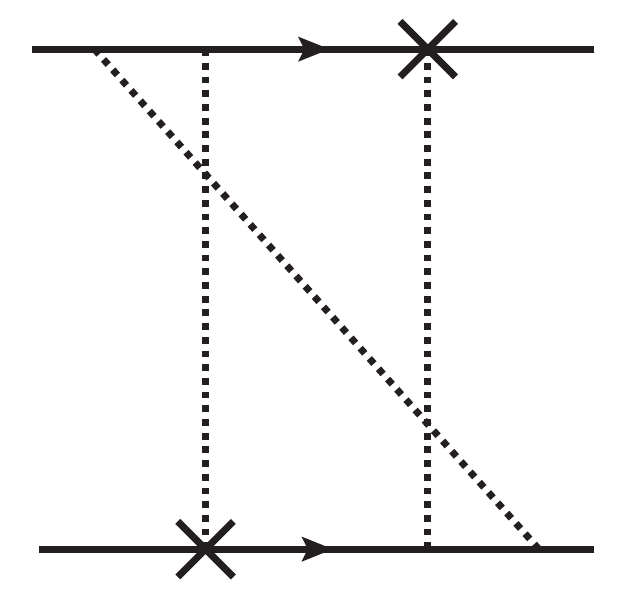}
	\put(-45,-10){(3e)}
	\includegraphics[width=0.25\columnwidth]{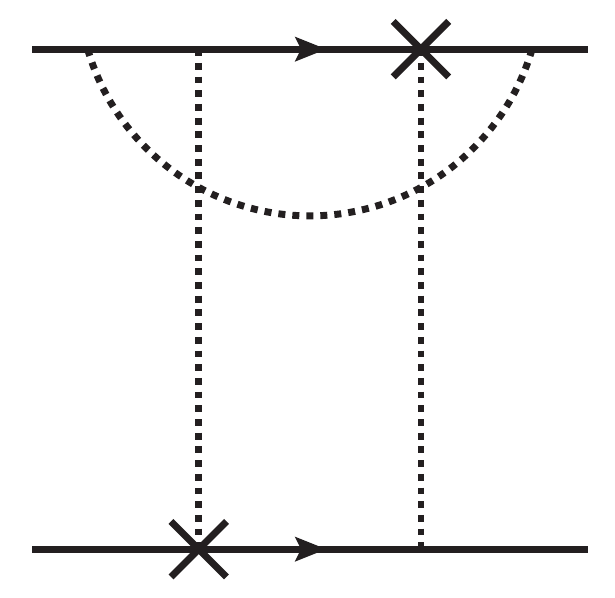}
	\put(-45,-10){(3f)}
	\includegraphics[width=0.25\columnwidth]{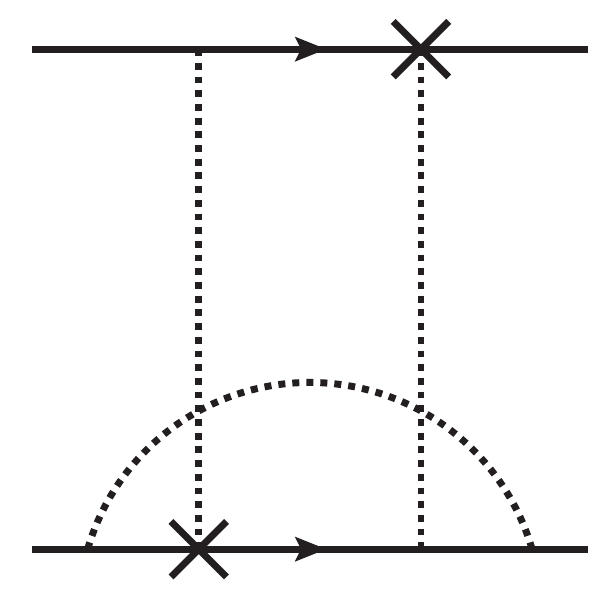}
	\put(-45,-10){(3g)}
	\vspace{1ex}
	\includegraphics[width=0.25\columnwidth]{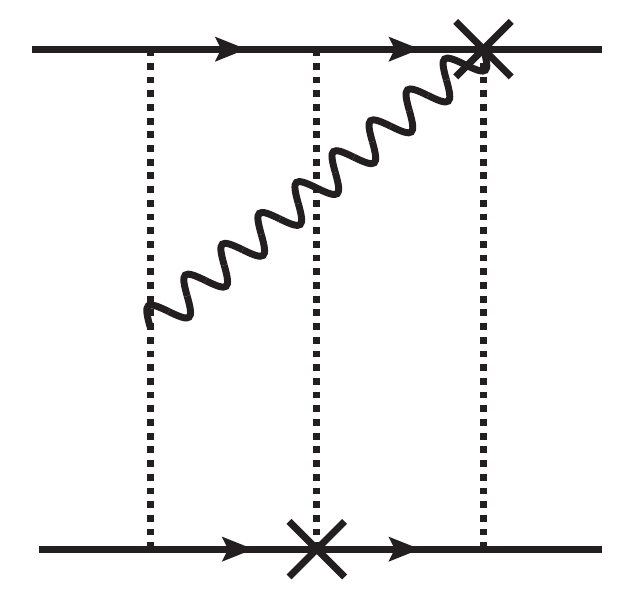}
	\put(-45,-10){(4a)}
	\includegraphics[width=0.25\columnwidth]{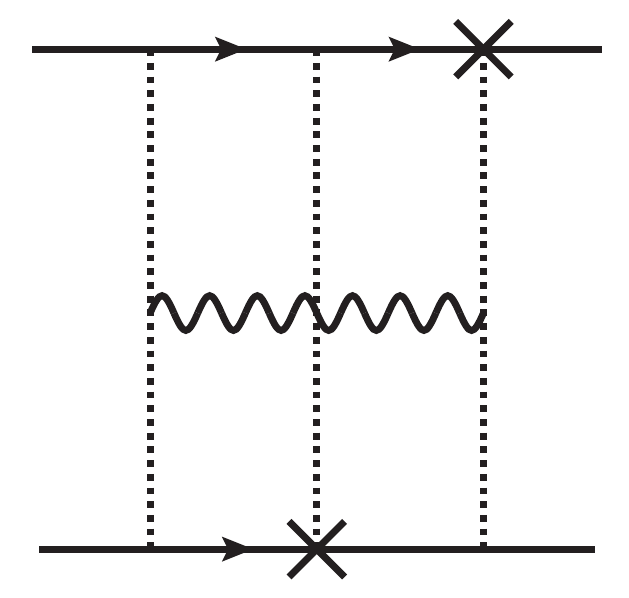} 
	\put(-45,-10){(4b)}
	\includegraphics[width=0.25\columnwidth]{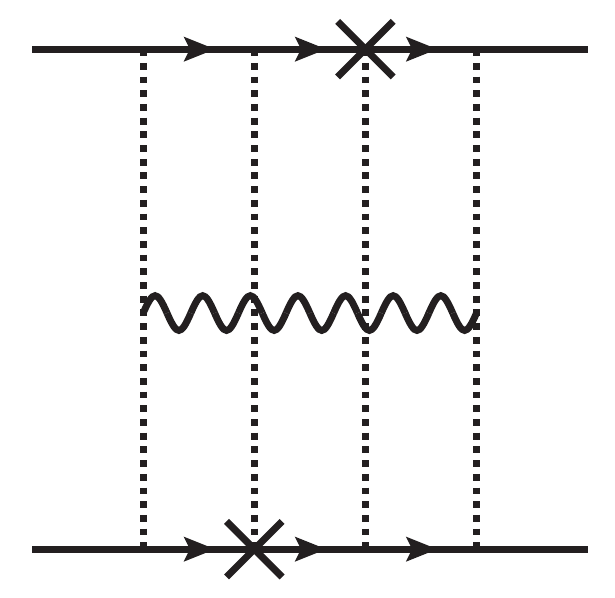} 
	\put(-45,-10){(5a)} 
\caption{Connected diagrams with one ultrasoft gluon exchange contributing to $E_{\rm US}^{(1,1)}$. Left-right mirror graphs are understood. As explained in the 
text, the diagrams are in one-to-one correspondence, and in fact equal, to the 
ones of \fig{Vx0Diags}.
\label{V11Diags}}     
\end{figure}

In \rcite{Peset:2015vvi} we obtained the NLO soft color-singlet contribution to \eqss{E1}{EL211}.
In contrast, the ultrasoft contribution to the NLO quasi-static energies cannot 
be calculated within conventional perturbation theory.
As the ultrasoft scale is only generated dynamically, it requires the 
resummation of LO potential (Coulomb) interactions.

In the Wilson loop calculation the ultrasoft scale $\Delta 
V\sim C_A \al / r$ is generated by the ratio of the resummed (exponential) LO expressions for the rectangular 
color-octet Wilson loop in the numerator and the color-singlet Wilson loop in 
the denominator in the definition of the quasi-static energies in 
\eqss{E1}{EL211}, cf. \eq{WLcorrelatorDef}.
This leads to a factor $\exp(-i \Delta V t)$, where $t$ is the (positive) time 
interval spanned by a (ultrasoft) gluon exchange between the quarks, in the 
otherwise perturbative calculation.
As will be explained below this exponential factor can be effectively treated 
as a position-space quark-antiquark pair propagator in our diagrammatic 
approach to the ultrasoft computation,  similar to the singlet and octet 
(bound-state) propagators in pNRQCD~\cite{Pineda:1997bj,Brambilla:1999xf}.

To achieve a clean and systematic separation from the soft contribution within 
dimensional regularization, we follow the method of 
regions~\cite{Beneke:1997zp}.
That is, we assign a definite energy and momentum scaling to the degrees of 
freedom involved in the ultrasoft calculation and consistently expand according 
to the power counting.
In our case the relevant degrees of freedom are the ultrasoft gluon with 
four-momentum $q^\mu \sim \Delta V$ and potential gluons with energy 
$k_0 \sim \Delta V$ and three-momentum $\bk \sim 1/r \gg \Delta V$.
The Wilson lines themselves can be interpreted as static (anti)quarks 
interacting with both types of gluons.
Modes with soft energies $\sim 1/r$ are insensitive to the ultrasoft scale 
and therefore decouple from the ultrasoft modes. They are not needed for the 
ultrasoft contribution to the NLO quasi-static energies.%
\footnote{At higher orders there can be mixed contributions with soft and 
ultrasoft modes.}
In this calculation only one ultrasoft gluon and a (infinite) number of 
potential gluons are involved. The couplings of the ultrasoft gluon yield 
a factor $\al$ relative to the LO, whereas any additional potential gluon 
exchange does not change the power counting. 

To compute the correlator $\langle\langle {\bf E}_a^i(t) {\bf 
E}_b^j(0)\rangle\rangle_c$ with two chromo-electric field operators ($a,b 
\in \{ 1,2 \}$) we use a (non-standard) diagrammatic approach.
The relevant (`connected') diagrams for the ultrasoft part of the quasi-static 
energies in \eqss{E1}{EL211} are displayed in \figs{Vx0Diags}{V11Diags}.
Vertical lines represent potential gluons, which do not propagate in time 
(after expansion in the ultrasoft momentum). Potential gluon interactions that 
will be resummed are not displayed. We will come back to this later.
The ultrasoft gluon is depicted as horizontal, diagonal or curvy line.
Before we comment on the selection criteria for these diagrams, we describe how 
to construct and evaluate them.

The procedure is as follows.
We first locate the vertices of the ultrasoft and potential gluons at 
specific times. 
As a consequence there is a priori no energy conservation at these vertices.
The two vertices for the chromo-electric field operator 
insertions on the static quark (Wilson) lines are fixed at times $0$ and $t$, 
respectively. The time coordinates of the other vertices have to be integrated 
over.
The energies in the potential gluon propagators are expanded out and do not 
contribute at NLO.%
\footnote{Strictly speaking, there are higher order terms in the potential 
energy expansion contributing in the Feynman gauge diagrams ($3e$), ($3f$) and 
($3g$), but they cancel out in the sum.}
The integration over these energies can therefore be carried out immediately 
yielding delta functions of time intervals.
Many of the time integrations can now be performed trivially.
The result is that vertices connected by potential gluons are all located at 
the same time.
This has been indicated already in our graphs in \figs{Vx0Diags}{V11Diags}, 
where time proceeds from left to right.

\begin{figure}[t]
\begin{center}      
\includegraphics[width=\columnwidth]{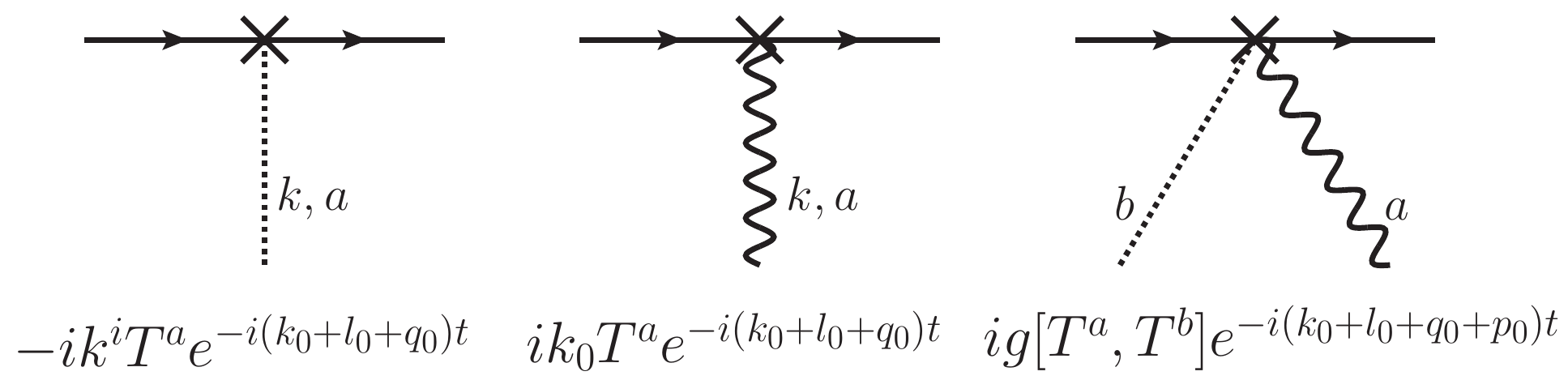}
\caption{
Feynman rules for an ${\bf E}^i(t)$ operator insertion (denoted by a cross in 
the diagram) on a static quark (Wilson) line. 
Dotted and wavy lines represent $A_0$ and $\bf A$ gluons, respectively.
All momenta ($k,l,q,p$) are incoming. 
Note that, in contrast to the usual convention for Wilson lines, we treat the 
antiquark (lower Wilson line) as a particle living in the anti-representation 
of $SU(3)$, i.e. the fermion flow (little arrows) of the antiquark is the same 
as for the quark. The corresponding Feynman rules for the antiquark are then 
obtained by replacing $g\rightarrow -g$ and $T^a \rightarrow (T^a)^T$. 
\label{EFeynRules}}  
\end{center}
\end{figure}

The vertices for the chromo-electric operator insertions are the same as the 
ones given in \rcite{Peset:2015vvi} and shown in \fig{EFeynRules}.
For the Wilson line interaction of an $A^0$ gluon, the quark and gluon propagators, and the gluon self-interactions we can use standard NRQCD Feynman rules at leading order in $1/m$.
In particular the static quark propagator in position space is nothing else 
than a $\theta$-function of the time difference.
We emphasize however that, according to the method of regions, we have 
to expand the potential gluon propagators in the energy and ultrasoft momenta.

\begin{figure}[t]
\begin{center}      
\includegraphics[width=0.35\columnwidth]{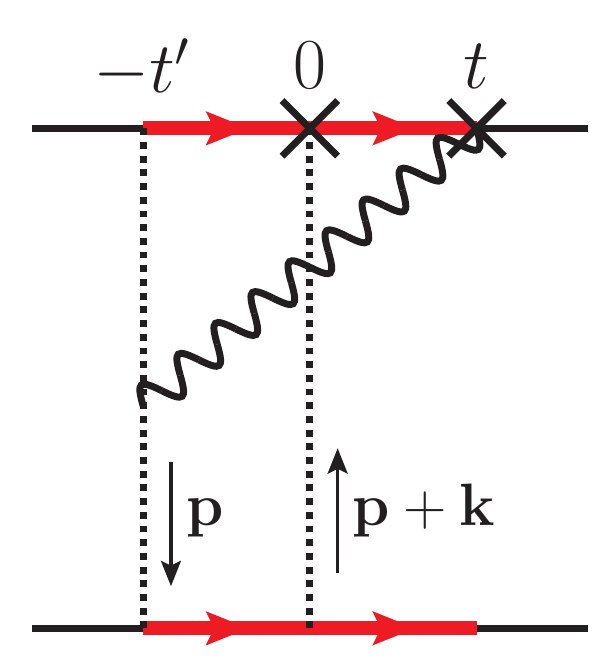} 
\caption{
Sample diagram ($3d$) contributing to $\delta E^{(2,0)}_{US}$ and $\delta 
E^{(1,0)}_{US}$, with time labels on the upper Wilson line vertices and 
indicated potential three-momentum flow. The thick red lines mark the time 
interval, where the static quark pair is in a color octet state.
\label{Diag3dEx}}  
\end{center}
\end{figure}

The way we implement the resummation of potential gluon interactions, which 
actually generates the ultrasoft scale, in our diagrammatic calculation is 
probably best explained at a concrete example.
Let us consider diagram ($3d$) in \fig{Vx0Diags}.
In \fig{Diag3dEx} we have depicted it with time labels for the vertices on the 
upper Wilson line and indicated a possible choice for the potential 
three-momentum flow. The momentum $\bk$ is the Fourier conjugate to the 
distance $\br$.
After expansion and  integration over the energies associated with the 
potential propagators the two vertices on the lower Wilson line are fixed at 
times $-t'$ and $0$, respectively.  
We have highlighted the time interval, where the static quark-antiquark system 
is in a color octet state, because the ultrasoft gluon has been radiated off, 
by thick red static quark (Wilson) lines.
During this time interval it is crucial to take into account the exchange of an 
infinite number of potential Coulomb gluons, between the two Wilson lines.
The resummation of such Coulomb interactions yields the LO expression for the 
average of a rectangular Wilson loop with extensions $t_2-t_1\gg r$ in the color 
octet configuration, see e.g. \rcite{Brambilla:1999xf}:
\begin{align}
&\hspace*{-0.5 cm}
\raisebox{-8mm}{\includegraphics[width=0.46\textwidth]{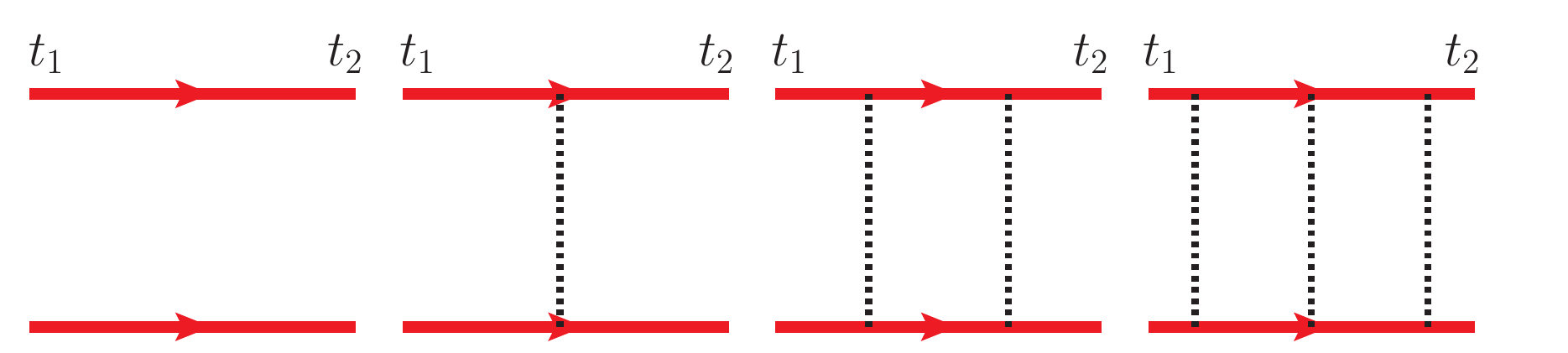}}
\put(-182,-3){+}
\put(-126,-3){+}
\put(-69,-3){+}
\put(-14,-3){+\;\ldots }
\nn\\
& = \theta(t_2\!-\!t_1)\, e^{-i V_o^{(0)} (t_2-t_1)}. 
\end{align}
The analogous expression with the octet static potential $V_o^{(0)}(r)$ replaced by 
the singlet static potential $V^{(0)}(r)$ holds for the color singlet configuration.
Outside the red marked time interval in \fig{Diag3dEx} this exponential however 
exactly cancels with the singlet Wilson loop average in the denominator of 
$\langle\langle {\bf E}_1^i(t) {\bf E}_1^j(0)\rangle\rangle$, 
whereas inside the octet time interval we can assign a resummed `propagator', 
$\theta(t_2-t_1) \exp[-i \Delta V (t_2-t_1)]$, to the static quark pair.

In Feynman gauge we thus have
\begin{align}
&\lim_{T\rightarrow \infty}\int_0^{T} \df t\; g_B^2\; t^n\;
\langle\langle {\bf E}_1^i(t) {\bf E}_1^j(0)\rangle\rangle \big|^{3d}_{\rm US}\nn\\
& =
2\,g_B^6 \bigg(\frac{C_A^2 C_F}{2} - C_A C_F^2 \bigg)
\int \!\! \frac{\df^Dq}{(2 \pi)^D} 
\int \!\! \frac{\df^{d}p}{(2 \pi)^d}
\int \!\! \frac{\df^{d}k}{(2 \pi)^d}
\nn\\
& \quad\times  e^{-i\bk \br} \;\frac{q_0}{q^2}\,\frac{ p^i (p^j+k^j)}{\bp^2 
(\bp+\bk)^4} \; 
\bigg( \int_0^\infty \!\! \df t'\; e^{-i (q_0 + \Delta V) t'}  
\bigg) \nn\\
&\quad\times\bigg( \int_0^\infty \!\! \df t \; t^n\; e^{-i (q_0 + \Delta V) t} 
\bigg)  \\ 
& = - 2\,g_B^6 \bigg(\frac{C_A^2 C_F}{2} - C_A C_F^2 \bigg)\,
\int \!\! \frac{\df^{d}p}{(2 \pi)^d}\,  \frac{p^i}{\bp^2} \, e^{-i\bp \br} \nn\\
&\quad\times\int \!\! \frac{\df^{d}k}{(2 \pi)^d}\,  \frac{k^j}{\bk^4} \, e^{-i\bk \br} \int \!\! \frac{\df^Dq}{(2 \pi)^D} \; \frac{q_0}{q^2}\bigg( \frac{i}{-q_0-\Delta V +i0} \bigg) \,
\nn \\ \label{Calc3d2}
&\quad  \times
\Bigg[ \Gamma(1+n)\bigg( \frac{i}{-q_0-\Delta V +i0}\bigg)^{1+n} \Bigg] \\
& =
- g_B^6 \bigg(\frac{C_A^2 C_F}{2} - C_A C_F^2 \bigg)\, 2^{4-2 D} \pi ^{1-\frac{3 
D}{2}} \, \frac{(-i)^{n+1}}{n+1} \nn\\
&\quad\times\, r^i r^j \, r^{4-2 D} \Delta V^{D-n-3} \, 
\sin \bigg(\frac{\pi  D}{2}\bigg)  \nn\\
&\quad \times \Gamma \bigg(3-\frac{D}{2}\bigg) \Gamma (D-4) \Gamma (D-1)  
\Gamma (3+n-D)
\,,
\label{Calc3d3}
\end{align}
where $D=d+1=4+2\eps$, $q$ denotes the ultrasoft four-momentum, $g_B$ is the bare version of the coupling, and we have 
multiplied by $2$ to add the left-right mirror graph exploiting time reversal 
symmetry.
We only kept the leading order in the ultrasoft momentum expansion. Higher 
orders would come with higher powers of $\al$ and are beyond the 
order we are interested in. For the same reason, we only need the LO expression
\begin{align}
\Delta V = \frac{g_B^2}{8} C_A \pi ^{\frac{1}{2}-\frac{D}{2}} r^{3-D} 
\Gamma \left(\frac{D-3}{2}\right) + \ord(g_B^4)
\label{DeltaVLO}
\end{align}
in $D$ dimensions and we can replace the bare coupling $g_B$ by the 
renormalized $g$ as $g_B=g+{\cal O}(g^3)$.

A few practical comments on this example are in order:
After performing the last two time integrations we have written \eq{Calc3d2} in 
a suggestive form in terms of momentum space building blocks.
The analogous form can be directly obtained for all diagrams in 
\figs{Vx0Diags}{V11Diags} by assigning the expression in square brackets to the 
momentum space quark-pair `propagator' between the two crossed vertices and 
\mbox{$i/(-q_0-\Delta V)$} to the other quark-pair `propagators' between an uncrossed 
and a crossed vertex, and using momentum space Feynman rules for the vertices 
and gluon propagators otherwise.

In Coulomb gauge all diagrams with an $A^0$ ultrasoft gluon vanish, because the 
$\bq$ integral is scaleless after the ultrasoft momentum expansion.
To change from Feynman to Coulomb gauge in the diagrams with an $\bA$ ultrasoft gluon we 
only have to replace $\delta^{ij} \to \delta^{ij}  - q^i q^j/\bq^2$ in the 
numerator of the $\bA$ gluon propagator.
It is easy to see that, due to the ultrasoft momentum expansion, this will result in nothing more than an additional 
overall factor $(D-2)/(D-1)$ for the Coulomb gauge diagrams.
In other words, due to gauge invariance, the sum of all diagrams with an 
ultrasoft $A^0$ gluon must be equal to $1/(D-2)$ times the sum of all diagrams 
with an ultrasoft $\bA$ gluon.
Note that in order to check this explicitly, it is crucial to use the 
$D$-dimensional expression for $\Delta V$ in \eq{DeltaVLO}.

Also, from our calculation in \eqs{Calc3d2}{Calc3d3} and the Feynman rules in 
\fig{EFeynRules} it should be clear that the diagrams ($3d$) in 
\figs{Vx0Diags}{V11Diags} yield exactly the same result and in fact all diagrams 
in \fig{Vx0Diags} equal their counterpart in \fig{V11Diags}.
As we will see below this is related to Poincar\'e invariance.
We therefore conveniently define the quantity
\begin{align}
{\cal W}_n^{ij} &= \lim_{T \rightarrow \infty}\int_0^T dt\; t^n\, \langle\langle 
g{\bf E}_1^i(t)g {\bf E}_1^j(0)\rangle\rangle_c \big|_{\rm US}\nn\\
&= \lim_{T \rightarrow \infty}\int_0^T dt\, t^n\, \langle\langle g{\bf 
E}_1^i(t)g {\bf E}_{2}^j(0)\rangle\rangle_c\big|_{\rm US} \;.
\label{Wdef}
\end{align}
with $n=1,2$ for the $1/m$ and $1/m^2$ quasi-static energies, respectively.
We can then write the ultrasoft contributions to the SI 
quasi-static energies in \eqss{E1}{EL211} as
\begin{align}	
\delta E_{\rm US}^{(1,0)}(r)&=-\frac{1}{2}\,\delta^{ij}\,{\cal W}_1^{ij} 
\,,\label{VUSdef10}\\
\delta E_{\bp^2,\rm US}^{(2,0)}(r)&=\frac{i}{2}\,{\hat \br^i}{\hat \br^j} 
\,{\cal W}_2^{ij} \,,\label{VUSdef20p}\\
\delta E_{\bL^2,\rm US}^{(2,0)}(r)&=\frac{i}{2(d-1)}\left(\delta^{ij}-d\,{ 
\hat \br^i }{\hat \br^j }\right){\cal W}_2^{ij} \,,\label{VUSdef20L}\\
\delta E_{\bp^2,\rm US}^{(1,1)}(r)&=i\,{\hat \br^i }{\hat \br^j 
}\,{\cal W}_2^{ij}\,,\label{VUSdef11p}\\
\delta 
E_{\bL^2,\rm US}^{(1,1)}(r)&=\frac{i}{d-1}\left(\delta^{ij}-d\,{\hat \br^i 
}{\hat \br^j }\right){\cal W}_2^{ij} \,.\label{VUSdef11L}
\end{align}
From these equations we can directly read off
\begin{align}
\delta E_{\bL^2,\rm US}^{(1,1)}(r) &= 2\, \delta E_{\bL^2,\rm US}^{(2,0)}(r)\,, \\
\delta E_{\bp^2,\rm US}^{(1,1)}(r) &= 2\, \delta E_{\bp^2,\rm US}^{(2,0)}(r)\, .
\end{align}
This verifies Poincar\'e invariance, see 
\rcite{Barchielli:1988zp}. It requires that, given that there is no ultrasoft contribution at $\ord(\al^2/m^0)$,  the $\ord(\al^2/m^2)$ ultrasoft part of the total energy of the quark-antiquark system must be proportional to  $1/m_r^2 = 1/m_1^2 +1/m_2^2 +2/(m_1 m_2)$.
The corresponding relations for the soft potentials at the 
same order were checked in \rcite{Peset:2015vvi}.

From the sum of all relevant ultrasoft one-loop diagrams we obtain the total 
bare result
\begin{align}
&{\cal W}_n^{ij} =
-\frac{i^{1-n}}{2 
  } \,g_B^2\,  C_F\, (D-2) \, \pi ^{-D/2} \; \; 
\nn\\
& \quad \Gamma \bigg(\frac{D}{2}\bigg)  \,
\Delta V^{D-n-1}  \Gamma (n+1-D) \Bigg(\delta^{ij}  \nn \\
&\quad + \hat\br^i \hat\br^j \; \frac{(D-3)  
(D-n-1) \big(D^2-D (n+5)+n+2\big)}{ (n+1) (n+2)}   \Bigg)
\,,
\label{TotWn}
\end{align}
in both Coulomb and Feynman gauge with $\Delta V$ as given in \eq{DeltaVLO}.
We emphasize that the connected diagrams in \fig{Vx0Diags} or \fig{V11Diags} 
are not the only diagrams contributing to \eq{TotWn}.
In our definition a `connected' graph cannot be split into two (lower loop) 
Wilson loop diagrams by cutting the quark and the antiquark line at the 
same time, i.e. by a vertical cut through the diagram.
We also have to take into account disconnected (2-Wilson-line reducible) 
one-loop diagrams as well as (products of) tree-level diagrams from the 
$\langle\langle \bE_a^i(t) \rangle\rangle \langle\langle \bE_b^j(0) 
\rangle\rangle$ term in the definition 
of the connected Wilson loop average, \eq{ConnectedWLdef}.
Their net effect is however only to remove the terms proportional to $C_F^2$ 
and $C_F^3$ from the connected diagram contributions as explained in detail 
in the next section.
Keeping this in mind it is straightforward to check that the diagrams given in 
\figs{Vx0Diags}{V11Diags} give the complete $C_F$ contribution we are 
interested in:
Every additional diagram with one ultrasoft gluon exchange you can draw, is 
either disconnected or redundant, in the sense that it 
is one of the diagrams with additional Coulomb interactions, which 
we automatically take into account by our resummation.
The individual contributions ($\propto C_F$) to ${\cal W}_n^{ij}$ from each 
diagram in \figs{Vx0Diags}{V11Diags} are given in 
\app{FGresults} in Coulomb and Feynman gauge respectively.


We can now use  \eq{TotWn} together with the LO expression \eq{DeltaVLO} for $\Delta V$ in the ultrasoft contributions \eqss{VUSdef10}{VUSdef11p} and add the bare results for the corresponding potentials in \rcite{Peset:2015vvi} to obtain finite expressions for the quasi-static energies.
With the bare coupling
\begin{align}
g_B^2=g^2\left(1+\frac{g^2\nu^{2\epsilon}}{4\pi}\left(\frac{e^{\gamma_E}}{4\pi}\right)^{\epsilon}\frac{\beta_0}{4\pi}\frac{1}{ \epsilon}\right)
\end{align}
expressed in terms of the $\MS$ renormalized coupling $\al \equiv \al(\nu)\equiv\frac{g^2\nu^{2\epsilon}}{4\pi}$, we find
\begin{align}
\label{E10SD}
E^{(1,0)}(r)&=V^{(1,0)}(r)+\delta E_{\rm US}^{(1,0)}(r)= \\
&\hspace{-0.5cm}-\frac{\alpha^2 C_A C_F }{4 r^2}\bigg\{1+\frac{\alpha }{2 \pi }\bigg[\frac{8}{3} C_A \ln (C_A \alpha e^{\gamma_E })+\frac{47}{18}C_A\nn\\
&\hspace{-0.5cm}-\frac{49 }{18}T_f n_f+2 \beta_0 \ln (e^{\gamma_E } \nu  r) \bigg]\bigg\}, \nn\\
\label{EP2SD}
E_{\vp^2}^{(2,0)}(r)&=V_{\vp^2}^{(2,0)}(r)+\delta 
E_{\vp^2,\,\rm US}^{(2,0)}(r)=\nn\\
&\hspace{-0.5cm}-\frac{\alpha ^2 C_A C_F}{3 \pi  r}\bigg\{\ln (  C_A \alpha e^{\gamma_E })+\frac{1}{4}\bigg\},\\
\label{EL2SD}
E_{{\bf L}^2}^{(2,0)}(r)&=V_{{\bf L}^2}^{(2,0)}(r)+\delta 
E_{{\bf L}^2,\,\rm US}^{(2,0)}(r)=\nn\\
&\hspace{-0.5cm}-\frac{\alpha ^2 C_A C_F}{3 \pi  r}\bigg\{2 \ln (C_A \alpha e^{\gamma_E })-1\bigg\},\\
\label{EP11SD}
E_{\vp^2}^{(1,1)}(r)&=V_{\vp^2}^{(1,1)}(r)+\delta 
E_{\vp^2,\,\rm US}^{(1,1)}(r)=\nn
\\
&\hspace{-0.5cm}-\frac{\alpha  C_F}{r}\bigg\{1+\frac{\alpha }{3 \pi  }\bigg[C_A \bigg(2 \ln (C_A \alpha e^{\gamma_E })-\frac{3}{2} \bigg)\nn\\
&\hspace{-0.5cm}+\beta_0 \bigg(\frac{3}{2} \ln (e^{\gamma_E } \nu  r)+\frac{1}{2}\bigg)\bigg]\bigg\},\\
\label{EL11SD}
E_{{\bf L}^2}^{(1,1)}(r)&=
V_{{\bf L}^2}^{(1,1)}(r)+\delta E_{{\bf L}^2,\,\rm US}^{(1,1)}(r)=\nn\\
&\hspace{-0.5cm}\frac{\alpha  C_F}{2 r}\bigg\{1-\frac{\alpha }{3 \pi}\bigg[2C_A \bigg(4 \ln (C_A \alpha e^{\gamma_E })-1\bigg)\nn\\
&\hspace{-0.5cm}+ \beta_0 \bigg(\frac{1}{4}-\frac{3}{2} \ln (e^{\gamma_E } \nu  r)\bigg)\bigg]\bigg\},
\end{align}
after setting $\epsilon \to 0$.

\subsection{Cancellation of $C_F ^2$ and $C_F ^3$ terms}

As already mentioned in the previous section, the Feynman diagrams we depict in 
\fig{Vx0Diags} and \fig{V11Diags} are the connected
(2-Wilson-line irreducible) diagrams relevant for our ultrasoft calculation.
It is clear that most of these diagrams are accompanied by a number of  
disconnected diagrams, which differ from the connected diagrams only by the 
order of the vertices on the quark lines. For each connected diagram in 
\fig{Vx0Diags} and \fig{V11Diags}, let us define a diagram `family', which also 
includes its disconnected relatives.

Regarding the associated color factor, within each family the connected diagram 
contains a maximally non-Abelian term, i.e.  a term that is linear in the 
fundamental Casimir constant $C_F$. 
The color factor of the disconnected diagrams consists of terms with at least 
two powers of $C_F$.
Summing all diagrams of a family and subtracting the corresponding 
$\langle\langle \bE_a^i(t) \rangle\rangle \langle\langle \bE_b^j(0) 
\rangle\rangle$ term in the definition of the quasi-static energies according 
to \eq{ConnectedWLdef} and \eq{Wdef}, all terms $\propto C_F^n$ with $n \ge 2$ 
cancel out and only the maximally non-Abelian term survives.
This mechanism is very much reminiscent of the well-known non-Abelian 
exponentiation~\cite{Gatheral:1983cz,Frenkel:1984pz,Schroder:1999sg} of the static potential.
For the case of the quasi-static potentials/energies we are lacking however a 
general proof. We have therefore checked the cancellation of the $C_F^2$ and 
$C_F^3$ terms present in the diagrams with one ultrasoft loop explicitly for 
each family.
In the following we will explain how this works for two diagram families, whose 
connected graphs are displayed in \fig{Vx0Diags}. From this it should be 
clear how to show the cancellation also for the other families.
In particular, all statements in this section can be straightforwardly 
extended and are therefore equally valid also for the diagrams in 
\fig{V11Diags}, which contribute to $\delta E^{(1,1)}_{US}$.

Let us first introduce the notion of a color-stripped diagram.
We write e.g.
\begin{align}	
\text{\fig{Vx0Diags}}\; (2c) \;=\; \left(C_F^2 - \frac{C_A C_F}{2}  \right)
\raisebox{-4mm}{\includegraphics[height=0.05\textheight]{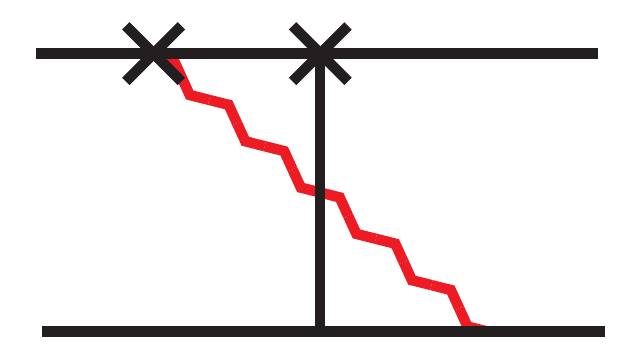}}\,.
\end{align}
The graph on the RHS symbolizes the original Feynman diagram with removed 
(fundamental, anti-fundamental or adjoint) color generators. The red zigzag line 
represents the ultrasoft gluon.

On the level of such color-stripped diagrams we have
\begin{align}
\raisebox{-4 mm}{\includegraphics[height=0.05\textheight]{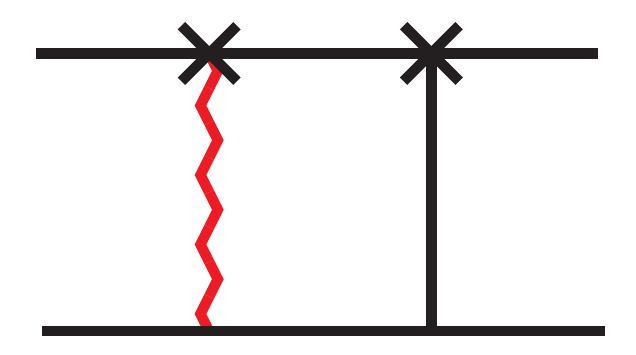}}
+
\raisebox{-4 mm}{\includegraphics[height=0.05\textheight]{diagrams/CS2c.pdf}}
=
\raisebox{-4 
mm}{\includegraphics[height=0.05\textheight]{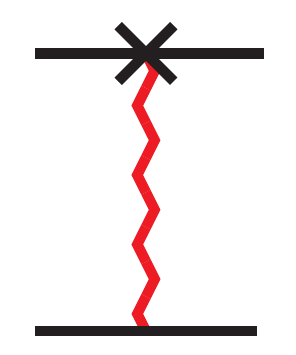}}
\times
\raisebox{-4 
mm}{\includegraphics[height=0.05\textheight]{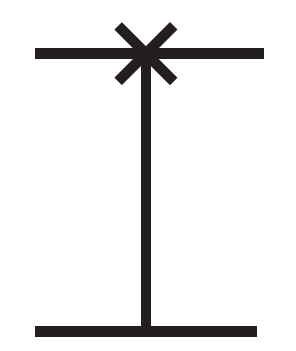}}
\,.
\end{align}
This can be seen easily as follows.
The only difference between the two graphs on the LHS is the lower quark 
propagator.
Labeling the vertex of the ultrasoft gluon with the lower quark line by the 
time 
$t'$ this propagator is $\theta(t'<t)$ and $\theta(t'>t)$ in the two diagrams, 
respectively.
After Coulomb resummation some of the $\theta$-functions actually come from the 
octet bound state propagators and all diagrams with an ultrasoft gluon have a 
factor $\exp(-i \Delta V |t'-t''|)$ in common, where in this case $t''=0$.
For the purpose of this section we can however effectively assign a
$\theta$-function to each static quark line in the color-stripped 
graphs.
In the sum of the two one-loop diagrams the $\theta$-functions drop out.
As we assume $t>0$ by definition, the upper `quark propagator' $\theta(t>0)$ 
can also be dropped.
Furthermore, by shifting the potential three-momentum $\bk$ we have (with $q$ 
being the ultrasoft four-momentum, $d=D-1$)
\begin{align}
\int\!\! \df^{d}q \int\!\! \df^{d}k\; \frac{e^{i \bk \br} q^i (q^j+k^j)}{
q^2\,(\bk+\bq)^2} &= \nn\\
&\hspace*{-2cm}\int\!\! \df^{d}q \; \frac{q^i}{q^2}\,  e^{-i \bq \br}  \times \int\!\! \df^{d}k\; \frac{k^j}{\bk^2} \,e^{i \bk \br}
\end{align}
in the integrand of the two one-loop diagrams.
Thus we are left with the product of the two tree-level diagrams on the RHS, which will eventually cancel with the corresponding term from $\langle\langle \bE_a^i(t) \rangle\rangle \langle\langle \bE_b^j(0) \rangle\rangle$).
Note that in this case the LO in the ultrasoft momentum expansion ($|\bq| \ll 
|\bk| \sim 1/r$) vanishes due to the antisymmetric $\bq$-integrand.
The NLO in this expansion is nonzero and represents the 
contribution at the order we are interested in.

With these preliminaries it is straightforward to show the cancellation 
of the $C_F^2$ and $C_F^3$ terms for most of the other diagram families, and  
we can tackle already the most complicated case, namely the family $3e$.
Without the maximally non-Abelian $C_A^2 C_F$ term the sum of the
diagrams in this family can be written as
\begin{align}
&-C_F^3\bigg(
\raisebox{-4 mm}{\includegraphics[height=0.05\textheight]{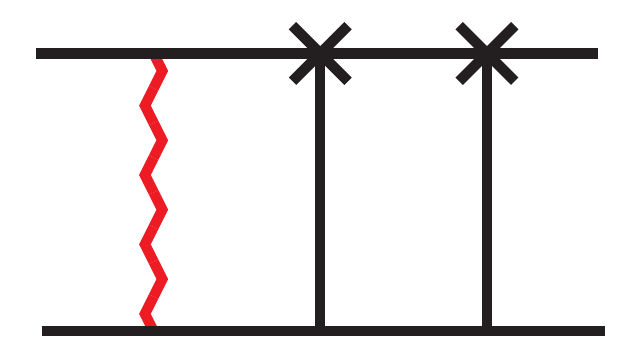}}
+
\raisebox{-4 mm}{\includegraphics[height=0.05\textheight]{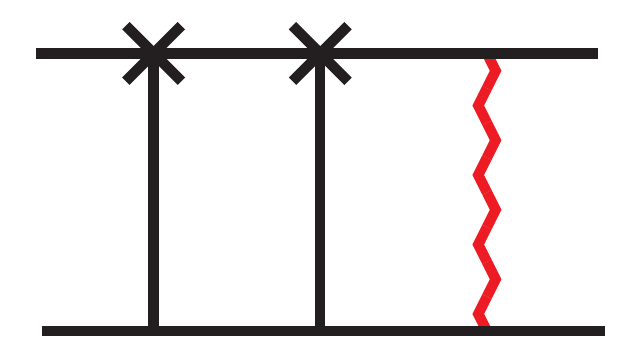}}
+
\raisebox{-4 mm}{\includegraphics[height=0.05\textheight]{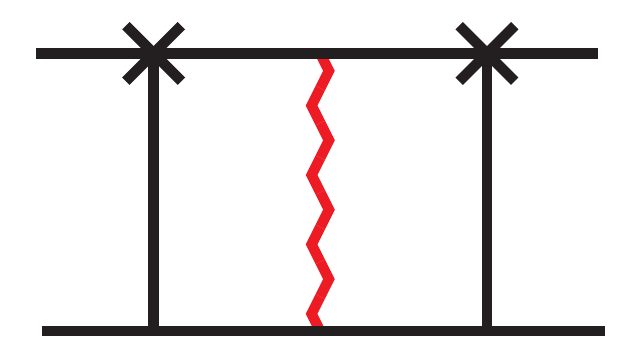}}
\bigg) \nn\\
&+\bigg(\frac{C_A C_F^2}{2}-C_F^3\bigg)\bigg(
\raisebox{-4 mm}{\includegraphics[height=0.05\textheight]{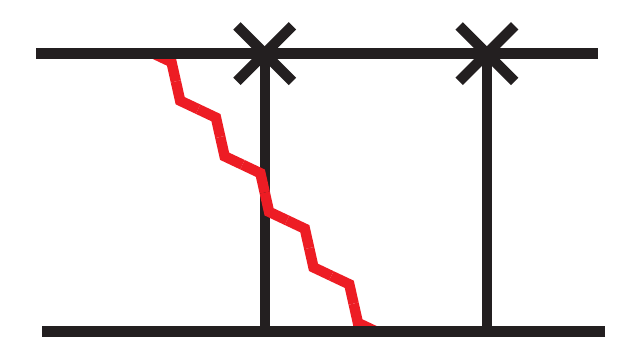}}
+
\raisebox{-4 mm}{\includegraphics[height=0.05\textheight]{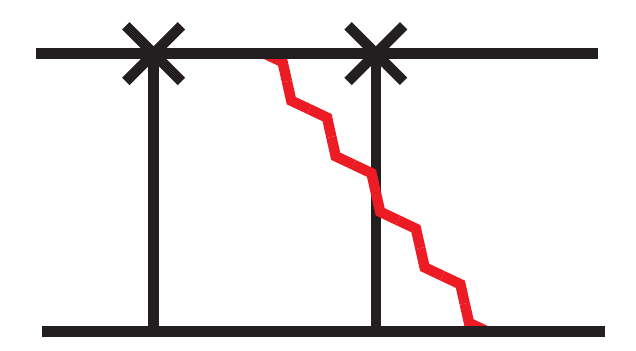}}
\nn\\
&+
\raisebox{-4 mm}{\includegraphics[height=0.05\textheight]{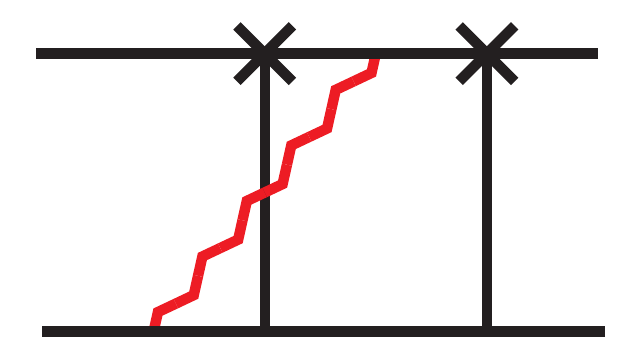}}
+
\raisebox{-4 mm}{\includegraphics[height=0.05\textheight]{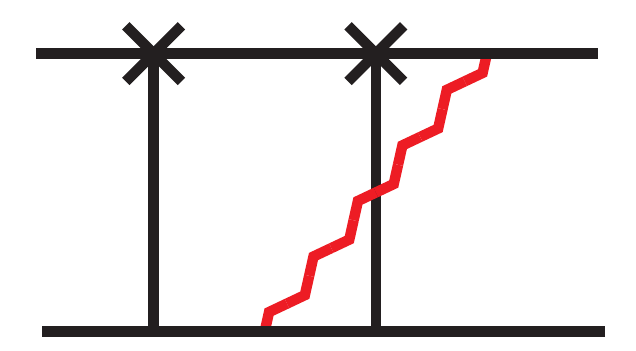}}
\bigg) \nn\\
&+\big(C_A C_F^2-C_F^3\big) \bigg(
\raisebox{-4 mm}{\includegraphics[height=0.05\textheight]{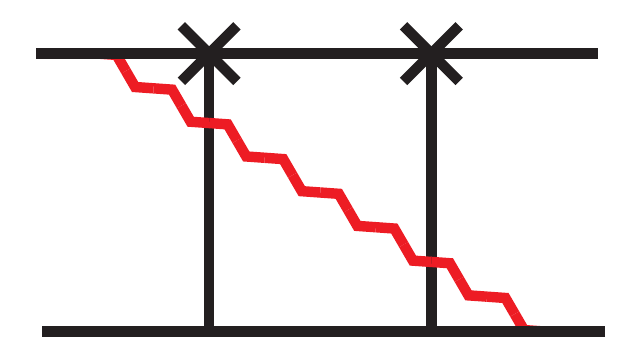}}
+
\raisebox{-4 mm}{\includegraphics[height=0.05\textheight]{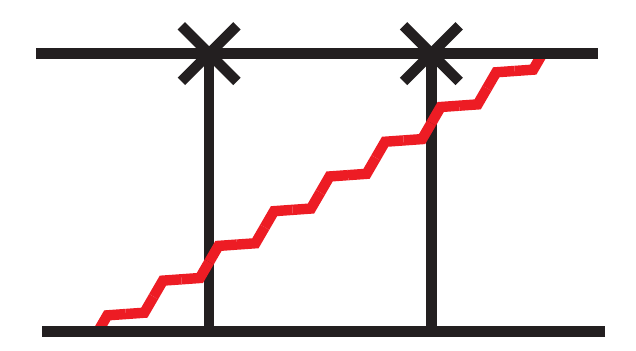}}
\bigg) \\
=& -C_F^3\bigg(
\raisebox{-4 mm}{\includegraphics[height=0.05\textheight]{diagrams/CS3e_1.pdf}}
+
\raisebox{-4 mm}{\includegraphics[height=0.05\textheight]{diagrams/CS3e_2.pdf}}
+
\raisebox{-4 mm}{\includegraphics[height=0.05\textheight]{diagrams/CS3e_3.pdf}}
\nn\\
&+
\raisebox{-4 mm}{\includegraphics[height=0.05\textheight]{diagrams/CS3e_4.pdf}}
+
\raisebox{-4 mm}{\includegraphics[height=0.05\textheight]{diagrams/CS3e_5.pdf}}
+
\raisebox{-4 mm}{\includegraphics[height=0.05\textheight]{diagrams/CS3e_6.pdf}}
\nn\\
&+
\raisebox{-4 mm}{\includegraphics[height=0.05\textheight]{diagrams/CS3e_7.pdf}}
+
\raisebox{-4 mm}{\includegraphics[height=0.05\textheight]{diagrams/CS3e_8.pdf}}
+
\raisebox{-4 mm}{\includegraphics[height=0.05\textheight]{diagrams/CS3e_9.pdf}}
\bigg) \nn\\
&+ \frac{C_A C_F^2}{2} \bigg(
\raisebox{-4 
mm}{\includegraphics[height=0.05\textheight]{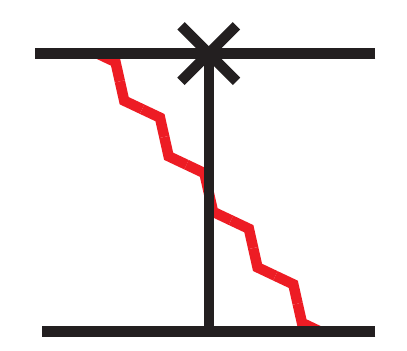}}
\!\!\times\!\! \raisebox{-4 
mm}{\includegraphics[height=0.05\textheight]{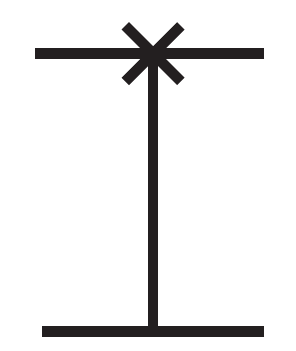}}
+
\raisebox{-4 
mm}{\includegraphics[height=0.05\textheight]{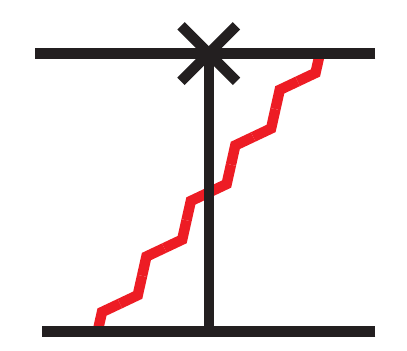}}
\!\!\times\!\! \raisebox{-4 
mm}{\includegraphics[height=0.05\textheight]{diagrams/CSsingno.pdf}}
\nn\\
&+
\raisebox{-4 
mm}{\includegraphics[height=0.05\textheight]{diagrams/CSsingno.pdf}}
\!\!\times\!\! \raisebox{-4 
mm}{\includegraphics[height=0.05\textheight]{diagrams/CSsingcross_1.pdf}}
+
\raisebox{-4 
mm}{\includegraphics[height=0.05\textheight]{diagrams/CSsingno.pdf}}
\!\!\times\!\! \raisebox{-4 
mm}{\includegraphics[height=0.05\textheight]{diagrams/CSsingcross_2.pdf}}
\bigg)\,.
\label{Exp3e}
\end{align}
Exploiting again the static quark line $\theta$-functions, we 
have used the equalities
\begin{align}
\raisebox{-4 mm}{\includegraphics[height=0.05\textheight]{diagrams/CS3e_4.pdf}}
+
\raisebox{-4 mm}{\includegraphics[height=0.05\textheight]{diagrams/CS3e_8.pdf}}
&=
\raisebox{-4 
mm}{\includegraphics[height=0.05\textheight]{diagrams/CSsingcross_1.pdf}}
\!\!\times\!\! \raisebox{-4 
mm}{\includegraphics[height=0.05\textheight]{diagrams/CSsingno.pdf}} 
\label{CSID1}
\;, \\
\raisebox{-4 mm}{\includegraphics[height=0.05\textheight]{diagrams/CS3e_5.pdf}}
+
\raisebox{-4 mm}{\includegraphics[height=0.05\textheight]{diagrams/CS3e_8.pdf}}
&=
\raisebox{-4 
mm}{\includegraphics[height=0.05\textheight]{diagrams/CSsingno.pdf}}
\!\!\times\!\! \raisebox{-4 
mm}{\includegraphics[height=0.05\textheight]{diagrams/CSsingcross_1.pdf}
} , 
\label{CSID2}
\end{align}
and their analogs for the graphs, where the ultrasoft gluon is up-down 
mirrored, in \eq{Exp3e} to write the 
$C_F^2$ term as a sum of products of one-loop and tree-level diagrams with a 
single insertion of an $\bE$-field operator.
Recall that in our notation the operator insertion vertex on the left is always 
located at time $0$ and the one on the right at time $t$ with $t>0$.
The two expressions on the RHS of \eq{CSID1} and \eq{CSID2} are therefore 
unequal.
The $C_F^2$ term of \eq{Exp3e} will be directly cancelled by the corresponding 
$C_F^2$ term from the
$\langle\langle \bE_1^i(t) \rangle\rangle \langle\langle \bE_1^j(0) 
\rangle\rangle$ part 
of the definition of the quasi-static energies. 

The cancellation of the $C_F^3$ term is a bit more subtle.
First we note that
\begin{align}
\raisebox{-4 mm}{\includegraphics[height=0.05\textheight]{diagrams/CS3e_3.pdf}}
 +
\raisebox{-4 mm}{\includegraphics[height=0.05\textheight]{diagrams/CS3e_6.pdf}}
+
\raisebox{-4 mm}{\includegraphics[height=0.05\textheight]{diagrams/CS3e_1.pdf}}\nn\\
 +
\raisebox{-4 mm}{\includegraphics[height=0.05\textheight]{diagrams/CS3e_4.pdf}}
=
\raisebox{-4 
mm}{\includegraphics[height=0.05\textheight]{diagrams/CSsingno.pdf}}
\!\!\times\!\! \raisebox{-4 
mm}{\includegraphics[height=0.05\textheight]{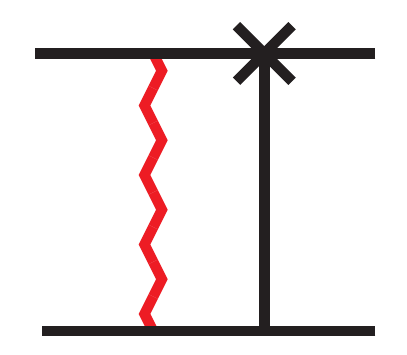}
}\,,
\label{CSID3}
\end{align}
by combining the $\theta$-functions from the upper and lower static quark lines.
Also, (after shifting a potential three-momentum) we directly have
\begin{align}
 \raisebox{-4 mm}{\includegraphics[height=0.05\textheight]{diagrams/CS3e_1.pdf}}
 =
\raisebox{-4 
mm}{\includegraphics[height=0.05\textheight]{diagrams/CSsingparallel_1.pdf}}
\!\!\times\!\! \raisebox{-4 
mm}{\includegraphics[height=0.05\textheight]{diagrams/CSsingno.pdf}
}\,.
\label{CSID4}
\end{align}
With the equalities in \eqss{CSID1}{CSID4} and their analogs for the mirror 
graphs we 
can write the $C_F^3$ term of \eq{Exp3e} in factorized form:
\begin{align}
& -\frac{C_F^3}{2}\Bigg[
\bigg(
\raisebox{-4 
mm}{\includegraphics[height=0.05\textheight]{diagrams/CSsingparallel_1.pdf}}
+
\raisebox{-4 
mm}{\includegraphics[height=0.05\textheight]{diagrams/CSsingcross_1.pdf}}
+
\raisebox{-4 
mm}{\includegraphics[height=0.05\textheight]{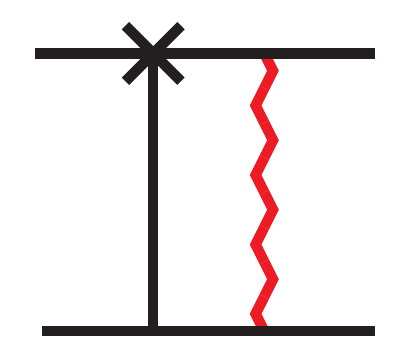}}\nn\\
&\qquad+
\raisebox{-4 
mm}{\includegraphics[height=0.05\textheight]{diagrams/CSsingcross_2.pdf}}
\bigg)
\times\!\! \raisebox{-4 
mm}{\includegraphics[height=0.05\textheight]{diagrams/CSsingno.pdf}
}
+
 \!\!
\raisebox{-4mm}{\includegraphics[height=0.05\textheight]{diagrams/CSsingno.pdf}}
\!\!\times 
\bigg(
\raisebox{-4 
mm}{\includegraphics[height=0.05\textheight]{diagrams/CSsingparallel_1.pdf}}\nn\\
&\qquad+
\raisebox{-4 
mm}{\includegraphics[height=0.05\textheight]{diagrams/CSsingcross_1.pdf}}
+
\raisebox{-4 
mm}{\includegraphics[height=0.05\textheight]{diagrams/CSsingparallel_2.pdf}}
+
\raisebox{-4 
mm}{\includegraphics[height=0.05\textheight]{diagrams/CSsingcross_2.pdf}}
\bigg)
\Bigg]\,.
\label{CF3term1}
\end{align}
Apart from the overall factor $1/2$ this is what one might naively expect for 
the $C_F^3$ term from $\langle\langle \bE_1^i(t) \rangle\rangle \langle\langle 
\bE_1^j(0) \rangle\rangle$ at NLO in the ultrasoft loop expansion.
On the first sight there therefore seems to be miscancellation of the $C_F^3$ 
terms.
Let us however take a closer look at \eq{CF3term1}. Similar to \eq{CSID3} the 
expressions in round brackets can be factorized further and we are left with
\begin{align}
 - C_F^3 
\raisebox{-4mm}{\includegraphics[height=0.05\textheight]{diagrams/CSsingno.pdf}}
\!\!\times\!\!
\raisebox{-4mm}{\includegraphics[height=0.05\textheight]{diagrams/CSsingno.pdf}}
\!\!\times\!\!
\raisebox{-4mm}{\includegraphics[height=0.05\textheight]{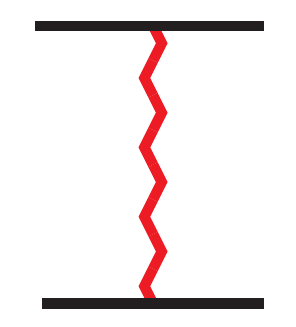}}
\,.
\label{CF3term2}
\end{align}
Now, recall that in the definition of $\langle\langle \bE_1(t) \bE_1(0) 
\rangle\rangle$ there is the singlet Wilson loop average in the denominator.
At LO we simply have $\langle  W_\Box\rangle$ = $\exp(-i V_s T_W)$, 
but at the NLO in the ultrasoft loop expansion we obtain
\begin{align}
C_F^3 
\raisebox{-4mm}{\includegraphics[height=0.05\textheight]
{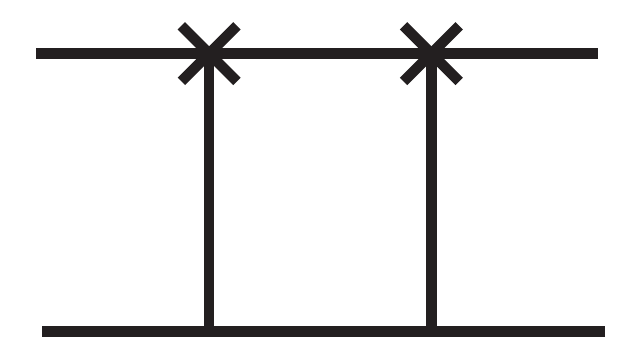}}
\!\!\times\!\!
\raisebox{-4mm}{\includegraphics[height=0.05\textheight]{diagrams/CSsingUS.pdf}}
=
 C_F^3 
\raisebox{-4mm}{\includegraphics[height=0.05\textheight]{diagrams/CSsingno.pdf}}
\!\!\times\!\!
\raisebox{-4mm}{\includegraphics[height=0.05\textheight]{diagrams/CSsingno.pdf}}
\!\!\times\!\!
\raisebox{-4mm}{\includegraphics[height=0.05\textheight]{diagrams/CSsingUS.pdf}}
\label{CF3term3}
\end{align}
as an additional contribution to the quasi-static energies.
Thus the total $C_F^3$ term from the $\langle\langle \bE_1^i(t) 
\bE_1^j(0)\rangle\rangle$ vanishes.
It is easy to see that the same happens separately for the $C_F^3$ term 
associated with $\langle\langle \bE_1^i(t)\rangle\rangle 
\langle\langle\bE_1^j(0)\rangle\rangle$.
There we have two times \eq{CF3term2} plus two times the RHS of \eq{CF3term3} 
from the expansion of the two denominators in  $\langle\langle 
\bE_1^i(t)\rangle\rangle 
\langle\langle\bE_1^j(0)\rangle\rangle$.
This completes the proof of the cancellation of the $C_F^2$ and $C_F^3$ terms 
related to family $3e$.

Analogous proofs hold for families $3f$ and $3g$.
Adding the contributions to ${\cal W}_n^{ij}$ from diagram families $3e$, $3f$ 
and $3g$ the LO terms 
in the ultrasoft momentum expansion, i.e. terms $\sim \al^3 \Delta V^{-1-n} 
r^{-4}$, cancel out, but a contribution $\sim 
\al^3 \Delta V^{1-n} r^{-2} \sim \al^{4-n} r^{n-3}$, which is the order 
of our interest, survives.
We note that the diagram families ($2c$, $3e$, $3f$, $3g$) discussed above 
actually vanish in Coulomb gauge.

For the families with an ultrasoft $\bA$ gluon ($3d$, $4a$, $4b$, $5a$) 
relevant also in Coulomb gauge the proofs are, in fact, even simpler than the 
ones presented here and straightforward to do with the above techniques.
We have thus shown `non-Abelian exponentiation' of the $1/m$ and $1/m^2$ 
quasi-static energies at one ultrasoft loop.


\section{Resummation of large logs}
\label{RG}

Our finite order results for the quasi-static energies in \eqss{E10SD}{EL11SD} depend on the factorization/renormalization scale $\nu$.
Setting $\nu \sim 1/r$ large logarithms associated with the soft scale are resummed in the running coupling $\al(\nu)$. For the complete resummation of large logarithms, also ultrasoft logarithms have to be resummed. In this section we discuss how this can be achieved.

The bare potential (in the Wilson loop scheme) can be written in terms of the renormalized potential and its counterterm as
\begin{align}
V_{s,B}=V_s+\delta V_s
\,.
\end{align}
The counterterm $\delta V_s$ for the potentials in terms of Wilson loops can be determined at leading order from the (IR) $1/\epsilon$ terms of the soft contribution computed in Ref.~\cite{Peset:2015vvi}. It can also be obtained (at leading order) from our ultrasoft calculation in the present paper. Indeed the $1/\epsilon$ terms of both computations match, which guarantees that we obtain finite results in \eqss{E10SD}{EL11SD}. At leading order the structure of the counterterm is (cf. \eq{deltaVsW} below)
\be
\delta V \sim \frac{1}{\epsilon} \al^n \times F({\bf r}^2)
\,,
\ee
where $n$ and $F$ depend on the specific potential $V$. This information by itself is not enough to obtain the leading-log (LL) running for each potential in the Wilson loop scheme. Still, this problem can be bypassed at LL, since EFT arguments tell us that the structure of the LO overall counterterm should be (note that $\nu$ here refers to $\nu_{us}$)
\be
\label{deltaVstr}
\delta V_s \sim \frac{1}{\epsilon} \al(\nu) V_A^2 \times G(V^{(0)};V_o^{(0)}),
\ee
where $G$ is a function of the singlet/octet static potentials, and $V_A=1$ at 
this order.

Let us now link this discussion to previous determinations of the counterterm $\delta V_s$ in the direct context of EFTs. If the counterterm is determined in an EFT calculation in terms of the 
Wilson coefficients of the EFT it is possible to resum the ultrasoft logarithms 
into the potentials by solving the associated renormalization group equations 
(RGEs).
Since our derivation of the quasi-static energies is based on the method of regions 
rather than on an EFT, we have a priori no RGEs at hand to resum such 
logarithms.  Using the EFT, the counterterm and the associated RGE for the next-to-leading-log (NLL) ultrasoft running of $V_s$ were obtained in \rcite{Pineda:2011aw}\footnote{Confirmation of the counterterm in the context of \mbox{vNRQCD}~\cite{Luke:1999kz} was 
obtained in \rcites{Hoang:2011gy,Hoang:2006ht} for the $\ord(1/m)$ and $\ord(1/m^2)$, 
potentials. 
The running of the static potential was computed at leading log (LL) in \rcite{Pineda:2000gza} and at NLL in \rcite{Brambilla:2009bi} and confirmed in \rcite{Pineda:2011db}.}
 (the LL ultrasoft running of $V_s$ was obtained in \rcite{Pineda:2001ra}).  
These expressions are suitable for the potentials in the on-shell or 
Feynman/Coulomb gauge off-shell matching schemes. However, as 
explained in \rcite{Peset:2015vvi}, the counterterm of \rcite{Pineda:2011aw} 
fails to cancel the divergences of the individual potentials in the Wilson loop 
matching scheme. 
Therefore we proposed a modified counterterm, that removes the 
divergences from the individual potentials at LO for that 
scheme~\cite{Peset:2015vvi}. Such a counterterm directly follows from the computation and the knowledge of the structure of the counterterm in \eq{deltaVstr}, as we discussed above.
We conjecture that the structure of this counterterm is preserved at NLO.
Adding the NLO piece according to \rcite{Pineda:2011aw} gives the  
following expression for the counterterm in the Wilson loop matching 
scheme ($W$):
\begin{align}
\nn
&\delta V_{s,W}
=
\Bigg(
{\bf r}^2(\Delta V)^3
-\frac{1}{2m_r^2}\left[{\bf p},\left[{\bf p},V^{(0)}_o\right]\right]\nn\\
&\quad+\frac{1}{2m_r^2}\left\{{\bf p}^2,\Delta V\right\}
+\frac{i}{2m_r^2}\left\{{\bf p}^i,\left\{{\bf p}^j,[{\bf p}^j,\Delta V r^i]\right\}\right\}
\nn\\
&\quad
\nn
+\frac{1}{2m_r}\Bigg[
(\Delta V)^2(3d-5)+4\Delta V\left( \left(r\frac{d}{dr}\Delta V\right)+\Delta V \right)\nn\\
&\quad+\left( \left(r\frac{d}{dr}\Delta V\right)+\Delta V \right)^2
\Bigg]
\Bigg)\nn
\\
&
\times
\Bigg[
\frac{1}{\epsilon}
C_F V_A^2
\bigg[
\frac{\al(\nu)}{3\pi}
-
\frac{\al^2(\nu)}{36\pi^2}
\bigg(\frac{10}{3}T_F n_f - C_A \bigg(\frac{47}{3}+2\pi^2 \bigg) \bigg)
\bigg]\nn\\
&\quad+
\frac{1}{\epsilon^2} C_F V_A^2
\frac{2}{3}\beta_0\frac{\al^2(\nu)}{(4\pi)^2}
\Bigg].
\label{deltaVsW}
\end{align}
At LO this is correct, because $\delta V_{s,W}$ correctly reproduces the 
UV divergence of the ultrasoft contribution to all quasi-static 
energies we have calculated.
At NLO and beyond it is only a conjecture, because in our approach we have no 
proper control on (potential) operator mixing effects in the RGE. We will 
nevertheless use the resulting expressions as an educated guess for the 
renormalized NLL potentials $V_{s,W}$. 
We hope that an EFT analysis will eventually settle this issue.

The RGE then reads
\begin{align}
\nu\frac{d}{d\nu} V_{s,\MS}
=
B_{V_s}
\,,
\end{align}
where
\begin{align}
\nn
B_{V_s}&=
C_F
\Biggl(
{\bf r}^2(\Delta V)^3
-\frac{1}{2m_r^2}\left[{\bf p},\left[{\bf p},V^{(0)}_o\right]\right]\nn\\
&+\frac{1}{2m_r^2}\left\{{\bf p}^2,\Delta V\right\}
+\frac{i}{2m_r^2}\left\{{\bf p}^i,\left\{{\bf p}^j,[{\bf p}^j,\Delta V r^i]\right\}\right\}
\nn\\
&
+\frac{1}{2m_r}
\left[ 4(\Delta V)^2
+4\Delta V\left( \left(r\frac{d}{dr}\Delta V\right)+\Delta V \right)
\right.\nn\\
&\left.+\left( \left(r\frac{d}{dr}\Delta V\right)+\Delta V \right)^2\right]
\Biggr)
\left[
-\frac{2\al(\nu)}{3\pi}\right.\nn\\
&\left.+
\frac{\al^2(\nu)}{9\pi^2}
\bigg(
\frac{10}{3}T_Fn_f -C_A \bigg(\frac{47}{3}+2\pi^2 \bigg)
\bigg)
+
{\cal O}(\al^3)
\right]
.
\end{align}
Here and in the following the Wilson loop scheme for the potentials is 
understood and we drop the corresponding subscript $W$ for brevity. We also 
set the pNRQCD Wilson coefficient $V_A=1$, which is consistent at LL~\cite{Pineda:2000gza} and NLL~\cite{Brambilla:2009bi}.

We write the running potential as
\begin{align}
V_{s,\rm RG}(r;\nu_{us}) & = V_{s}(r;\nu)+ \delta V_{s,\rm 
RG}(r;\nu,\nu_{us})\,,
\label{Vsrun}
\end{align}
with $\delta V_{s,\rm RG}(r;\nu,\nu_{us}\!=\!\nu) = 0$.
We now only consider the $1/m$ and the $1/m^2$ SI momentum-dependent potentials.
Solving the RGEs and truncating at the appropriate order in $\al$ we obtain
\begin{align}
\delta V^{(1,0)}_{\rm RG}(r;\nu,\nu_{us})&=2\left(\frac{C_A \alpha(\nu) }{2  
r}\right)^2\nn\\
&\hspace*{-2.2cm}\times
\left(
1+2\frac{\al(\nu)}{4\pi}\left(a_1+2\beta_0\ln(\nu e^{\gamma_E+\frac{1}{2}}r)\right)
\right)
F(\nu;\nu_{us})\,,
\\
\delta V^{(2,0)}_{{\bf p}^2,\rm RG}(r;\nu,\nu_{us})&=\nn\\
&\hspace*{-2.2cm}\frac{C_A \alpha(\nu) }{2  
r}
\left(
1+\frac{\al(\nu)}{4\pi}\left(a_1+2\beta_0\ln(\nu e^{\gamma_E+2}r)\right)
\right)
F(\nu;\nu_{us})\,,
\\
\delta V^{(2,0)}_{{\bf L}^2,\rm RG}(r;\nu,\nu_{us})&=\nn\\
&\hspace*{-2.2cm}\frac{C_A \alpha(\nu) }{  
r}
\left(
1+\frac{\al(\nu)}{4\pi}\left(a_1+2\beta_0\ln(\nu e^{\gamma_E-1}r)\right)
\right)
F(\nu;\nu_{us})\,,
\\
\delta V^{(1,1)}_{{\bf p}^2,\rm RG}(r;\nu,\nu_{us}) &=\nn\\
&\hspace*{-2.2cm}\frac{C_A \alpha(\nu) }{  
r}
\left(
1+\frac{\al(\nu)}{4\pi}\left(a_1+2\beta_0\ln(\nu e^{\gamma_E+2}r)\right)
\right)
F(\nu;\nu_{us})\,,
\\
\delta V^{(1,1)}_{{\bf L}^2,\rm RG}(r;\nu,\nu_{us})&=\nn\\
&\hspace*{-2.2cm}\frac{2C_A \alpha(\nu) }{  
r}
\left(
1+\frac{\al(\nu)}{4\pi}\left(a_1+2\beta_0\ln(\nu e^{\gamma_E-1}r)\right)
\right)
F(\nu;\nu_{us})\,,
\end{align}
where
\begin{align}
&F(\nu;\nu_{us})
=C_F
\frac{2\pi}{\beta_0}
\bigg\{
\frac{2}{3\pi}\ln\frac{\al(\nu_{us})}{\al(\nu)}
-\big(\al(\nu_{us})-\al(\nu)\big)\nn\\
&\times
\left(
\frac{8}{3}\frac{\beta_1}{\beta_0}\frac{1}{(4\pi)^2}-\frac{1}{27\pi^2}\left(C_A\left(47+6\pi^2\right)-10T_Fn_f\right)
\right)
\bigg\}
\,.
\end{align}

We computed the $\MS$ renormalized (Wilson loop) potentials at 
$\nu_{us}\!=\!\nu$ in \rcite{Peset:2015vvi}. For convenience we quote the 
results here:%
\footnote{Though not relevant for this work, we would like to point out a typo in the first line of Eq.~(5.24) of that reference, where $C_F\al/2$ should read $C_F\al/4$.}
\begin{align}
\label{Vren2W}
V^{(1,0)}(r;\nu)&=-\frac{C_FC_A\alpha^2(e^{-\gamma_E}/r)}{4r^2}\nn\\
&\hspace*{-1cm}\left\{1+\frac{\alpha}{\pi}\left(\frac{89}{36}C_A-\frac{49}{36}T_Fn_f+\frac{4}{3}C_A \ln\left(\nu r e^{\gamma_E}\right)\right)\right\},\\
V^{(2,0)}_{\vp^2}(r;\nu)&=-\frac{C_A C_F \alpha ^2 }{ \pi  r}\left(\frac{2}{3}+\frac{1}{3}\ln\left(r \nu  e^{\gamma_E}\right)\right) ,
\\
V^{(2,0)}_{{\bf L}^2}(r;\nu)&=\frac{C_A C_F \alpha^2 }{4 \pi  r}\left(\frac{11}{3}-\frac{8}{3}\ln\left(r \nu  e^{\gamma_E}\right)\right) ,
\\
V^{(1,1)}_{\vp^2}(r;\nu)&=-C_F \frac{\alpha(e^{-\gamma_E}/r) }{r}\nn\\
&\hspace*{-1cm}\left\{1+\frac{ \alpha }{ \pi }\left( \frac{23 }{18 }C_A-\frac{2}{9 }T_F n_f+\frac{2}{3}C_A \ln\left(\nu r e^{\gamma_E} \right)\right)\right\} ,\\
V^{(1,1)}_{{\bf L}^2}(r;\nu)&=\frac{C_F \alpha(e^{-\gamma_E}/r) }{2r}\nn\\
&\hspace*{-1cm}\left\{1+\frac{\alpha }{ \pi }\left(\frac{97 C_A}{36 }+\frac{1}{9 }T_F n_f -\frac{8 }{3 }C_A \ln\left(\nu r e^{\gamma_E} \right)\right)\right\}.	
\end{align}

Including the resummation of ultrasoft logarithms the total quasi-static 
energy then reads
\begin{align}
\label{ERG}
E_{s,\rm RG}(r) & = V_{s,\rm RG}(r;\nu_{us})+\delta E_{{\rm US}}(r;\nu_{us})\,,
\end{align}
and similarly for the individual quasi-static energies we discuss in this paper.
The corresponding $\MS$ renormalized ultrasoft parts are
\begin{align}
\label{Eren2WL}
\delta E^{(1,0)}_{\rm US}(r;\nu_{us})&=\nn\\
&\hspace*{-1cm}-\frac{ \alpha^2(\nu)\alpha(\nu_{us}) C_F C_A^2}{3 \pi  r^2}
\left\{
\ln 
\left(
\frac{C_A\,\alpha(\nu) }{\nu_{us}  r}\right)
-\frac{7}{8}
\right\},\\
\delta E^{(2,0)}_{{\bf p}^2,\rm US}(r;\nu_{us})&=\nn\\
&\hspace*{-1cm}
-\frac{  \alpha(\nu)\alpha(\nu_{us}) C_F C_A}{3 \pi  r}
\left\{
\ln 
\left(
\frac{C_A\,\alpha(\nu) }{\nu_{us}  r}\right)
-\frac{7}{4}
\right\},\\
\delta E^{(2,0)}_{{\bf L}^2,\rm US}(r;\nu_{us})&=\nn\\
&\hspace*{-1cm}
-\frac{  2\alpha(\nu)\alpha(\nu_{us}) C_F C_A}{3 \pi  r}
\left\{
\ln 
\left(
\frac{C_A\,\alpha(\nu) }{\nu_{us}  r}\right)
+\frac{7}{8}
\right\},\\
\delta E^{(1,1)}_{{\bf p}^2,\rm US}(r;\nu_{us})&=\nn\\
&\hspace*{-1cm}
-\frac{ 2 \alpha(\nu)\alpha(\nu_{us}) C_F C_A}{3 \pi  r}
\left\{
\ln 
\left(
\frac{C_A\,\alpha(\nu) }{\nu_{us}  r}\right)
-\frac{7}{4}
\right\},\\
\delta E^{(1,1)}_{{\bf L}^2,\rm US}(r;\nu_{us})&=\nn\\
&\hspace*{-1cm}
-\frac{ 4 \alpha(\nu)\alpha(\nu_{us}) C_F C_A}{3 \pi  r}
\left\{
\ln 
\left(
\frac{C_A\,\alpha(\nu) }{\nu_{us}  r}\right)
+\frac{7}{8}
\right\}
.
\label{Eren2W}
\end{align}
Formally, these ultrasoft contributions are independent of $\nu$, as in pNRQCD 
the soft scale has already been integrated out. 
In practice they do depend on $\nu$ once we allow for the running of 
$\alpha(\nu)$. 
This however affects the result only beyond the order to which the ultrasoft 
computation has been carried out. 
For a meaningful evaluation of the (RG improved) quasi-static energies $\nu \approx 1/r$ and $\nu_{us} \approx C_A \al/r$ must be chosen.

\section{Comparison with lattice data}
\label{Sec:latt}
 
Monte Carlo lattice simulations of the $1/m$ and $1/m^2$ SI 
momentum-dependent 
quasi-static energies have been performed over the 
years~\cite{Bali:1997am,Koma:2007jq,Koma:2009ws}. The data for 
the SI momentum-dependent $1/m^2$ contributions is usually 
displayed in terms of the following linear combinations of the quasi-static 
energies:%
\footnote{In the literature often the notation  $V_b$, $V_c$, etc. is 
used. We stick however to the notation $E_b$, $E_c$, etc. to distinguish the 
quasi-static energies studied here from the potentials that 
appear in the Schr\"odinger equation.}
\begin{align}
E_b(r)&= -\frac{2}{3} E_{\textbf L^2}^{(1,1)}(r) -E_{\vp^2}^{(1,1)}(r)
\label{Vbdef},\\
E_c(r)&= -E_{\textbf L^2}^{(1,1)}(r),\\
E_d(r)&= \frac{2}{3}E_{\textbf L^2}^{(2,0)}(r)+ E_{\vp^2}^{(2,0)}(r),\\
E_e(r)&= E_{\textbf L^2}^{(2,0)}(r).
\label{Vddef}
\end{align}
Therefore, we will directly confront our results with these quantities in the 
following section.
On the lattice side we will use the most recent results obtained in 
\rcites{Koma:2007jq,Koma:2009ws}, also for the comparison with the $1/m$ 
quasi-static energy.

\subsection{Short distances}
\label{Sec:lattshort}
Lattice data is very scarce at short distances. It is also worth 
mentioning that at short distances one can easily see finite discretization 
effects in the lattice data. On top of that, due to possible power-like 
divergences in the lattice simulation, we cannot compare with the lattice data 
in absolute values. We are thus forced to consider differences between
 quasi-static energies computed at different distances. We then choose to normalize 
our results to the lattice point at the shortest available distance for each 
observable.
This leaves us with at most three independent lattice points (in fact 
only two for $E_b$ and $E_c$) for distances smaller than $0.5\,r_0 
\approx 0.8$ GeV, which is already at the borderline for our weak coupling results to be reliable. 
Therefore, the whole discussion in this section will stay at the qualitative level. 

The analytic results we present in the plots of this section are based on 
\eqss{E10SD}{EL11SD}. For the resummation of the ultrasoft logarithms we will 
use the the results of \Sec{RG}. We will first define the different curves 
shown in these plots and discuss them in comparison with the available lattice 
data afterwards. The different colors of the lattice data points in our plots 
(with hardly visible error bars) indicate different $\beta$ values (blue: 
$\beta=5.85$, green: $\beta=6$, brown: $\beta=6.2$), see 
\rcites{Koma:2007jq,Koma:2009ws}.
For $E^{(1,0)}$, $E_b$ and $E_c$ \eqss{E10SD}{EL11SD} give the first two 
(LO and NLO) nontrivial terms, whereas for $E_d$ and $E_e$ they only yield the 
first (NLO) nontrivial term in the weak coupling expansion. 
The scale dependence of the coupling constant is uncancelled beyond NLO. 
We use this fact to grasp the dependence of our predictions on higher order 
corrections. 
On the one hand we produce curves with a fixed, $r$-independent 
factorization scale $\nu$. For these, we take as a reference scale the 
shortest distance of the available lattice points, and vary $\nu$ around this 
value.
For $E^{(1,0)}$, $E_d$ and $E_e$, we take $\nu=x/0.153990\;r_0^{-1}$ and 
for $E_b$ and $E_c$  we take $\nu=x/0.371627\;r_0^{-1}$.
Here and in the following the parameter $x$ always represents a number of 
order $1$.
On the other hand we also produce curves with $\nu=x/r$. 
The obvious motivation of this is to resum logarithms of $\nu 
r$.%
\footnote{Setting $\nu=1/r$ may introduce problems with renormalons (see the 
discussion in \rcite{Pineda:2002se}), which we do not explore here. The first 
reason is that a complete analysis of the renormalon structure of these 
potentials is missing (besides the trivial fact that $E^{(2,0)}_{\bf p^2}$ 
depends on the leading pole mass renormalon). Moreover, we are at low orders in 
the weak coupling expansion and indications of renormalons, even if existing, cannot be clearly distinguished from other uncertainties.} 
This is however not sufficient to resum all large logarithms, as there 
are also ultrasoft logarithms. 
In order to perform a complete resummation of logarithms, we have to use 
\eqss{Vsrun}{Eren2W} and count $\al(\nu) \sim \al(\nu_{us})$, so that the 
resummation kernel $F(\nu,\nu_{us}) \sim 1$.
With this counting the ${\cal O}(\al^2)$ term in the expression for the $1/m$ 
and the ${\cal O}(\al)$ term in the expression for the $1/m^2$ quasi-static 
energies, represents the results with LL accuracy.
Our complete expressions for the quasi-static energies according to \eq{ERG} 
represents an educated guess of
the NLL expression and we use it as an estimate of the next order.
The dependence on $\nu$ and $\nu_{us}$ 
cancels in the theoretical expression at NLL but not beyond. 
Again, this residual dependence may give hints on the size of higher 
order terms. We will generate lines with $\nu=x/r$ and $\nu_{us}=
x_{us}C_A\als(\nu)/r$, with $x \sim x_{us} \sim 1$.  
Note that the RG improved expressions reduce to the `fixed order' 
expressions in \eqss{E10SD}{EL11SD} when fixing $\nu=\nu_{us}$ in 
\eqss{Vsrun}{Eren2W}.

\begin{figure}[t]
\begin{center}      
\includegraphics[width=\columnwidth]{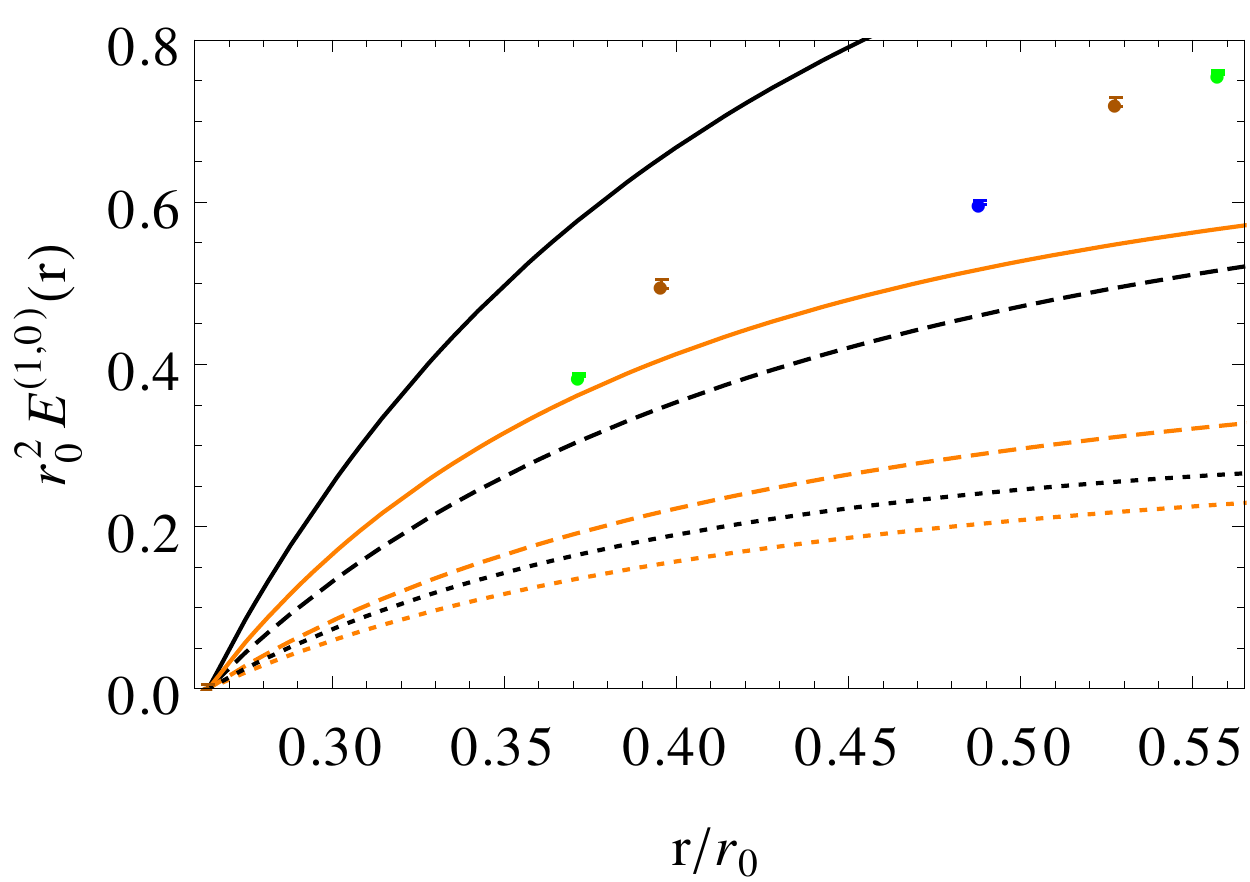}
\caption{Comparison to the lattice data for $E^{(1,0)}(r)$. The 
solid lines correspond to the LO (orange) and NLO results (black) at 
fixed scale $\nu=2\times 1/0.263821$ $r_0^{-1}$. 
The dashed lines correspond to the LO (orange) and NLO (black) 
setting $\nu=3/r$. The dotted lines correspond to the LL (orange) and 
NLL results (black) setting $\nu_{us}=2 C_A\alpha(\nu)/r$ and 
$\nu=3/r$. 
\label{FigV10SD}}
\end{center}
\end{figure}
We first consider the $1/m$ quasi-static energy at LO and NLO. 
We observe that the uncertainties are quite large.
This is reflected in a strong scale dependence, which could be 
expected as we are evaluating our result at very low scales and already 
the LO is ${\cal O}(\al^2)$. Even using the one-loop or two-loop expression for the running $\al$ 
produces sizable differences.
Nevertheless, it is comforting that we can find reasonable values 
of $\nu$ such that the curves match the lattice data quite well
($\nu \approx 3.4/0.264 \; r_0^{-1} \approx 5.2$ GeV for the NLO and $\nu 
\approx
1.6/0.264 $ $r_0^{-1} \approx 2.4$ GeV for the LO), even though the difference 
between LO and NLO is relatively large.
To illustrate this difference we show curves for the LO and NLO results 
with fixed $\nu=2\times 1/0.263821$ $r_0^{-1}\approx 3$ GeV in
\fig{FigV10SD}.  
We also show the corresponding curves with $\nu=3/r$.
They hardly differ from the choice $\nu=2/r$ and lie below the lattice points, 
but are still consistent with them within the expected uncertainties. 
The effect of setting $\nu \ge 2/r$ is 
making the series more convergent: 
The difference between LO and NLO becomes smaller compared to the 
fixed scale results as observed in \fig{FigV10SD}. 
However, $\nu \le 1/r$ is a too small scale and the perturbation 
series blows up (which may be due to renormalons).
A systematic resummation of all large logarithms requires setting 
$\nu_{us} \sim C_A\al(\nu)/r$ and $\nu \sim 1/r$. 
The additional resummation of ultrasoft logarithms further improves 
the convergence (LO and NLO curves are closer), but the curves also lie further below the lattice points, see \fig{FigV10SD}. 
Just like for the `fixed order predictions' with $r$-dependent factorization 
scale, we have to choose relatively large values of $\nu$ to avoid unphysical behavior.

\begin{figure}[t]
\begin{center}    
\includegraphics[width=\columnwidth]{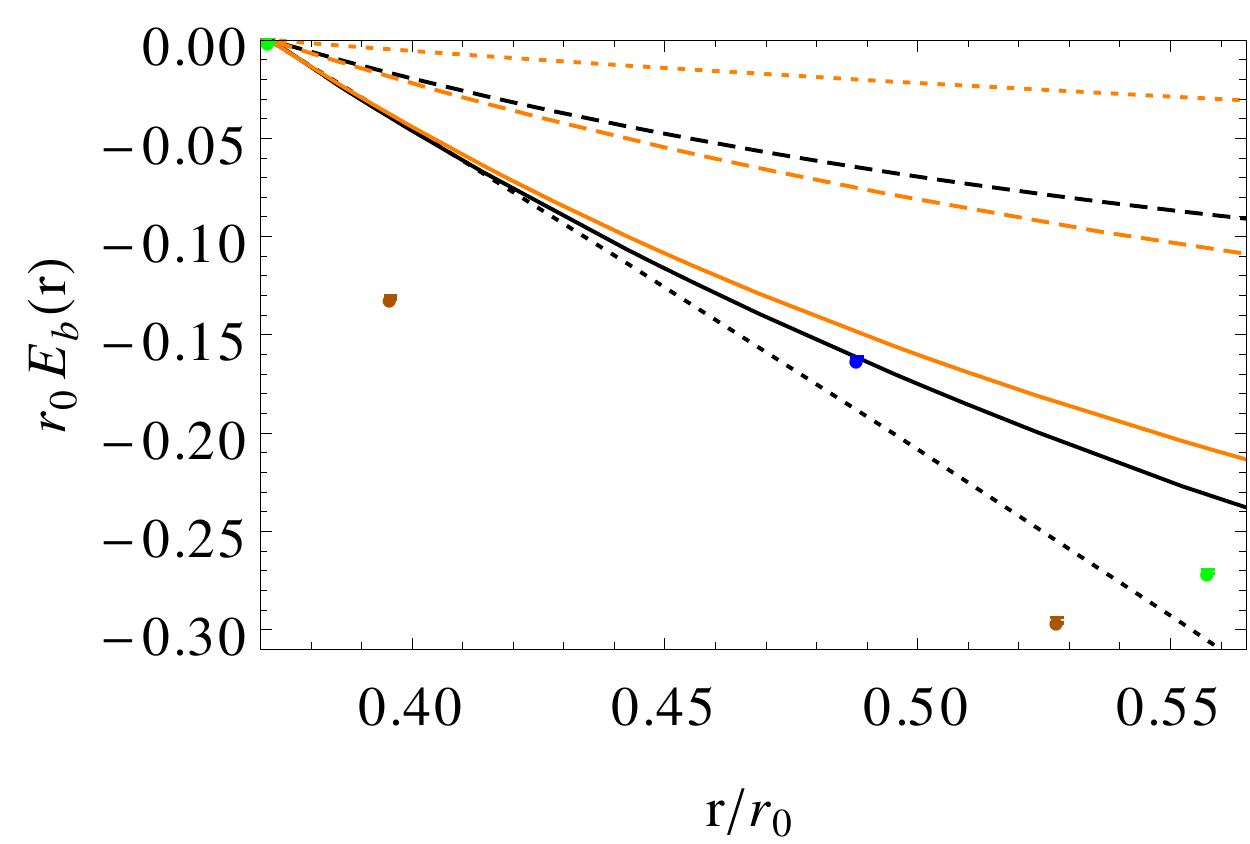}
\caption{Comparison to the lattice data for $E_b$. 
The solid lines correspond to the LO (orange) and NLO 
results (black) at fixed scale $\nu=2\times 1/0.371627$ $r_0^{-1}$. The 
dashed lines correspond to the LO (orange) and NLO results (black) 
setting $\nu=4/r$. The dotted lines correspond to the LL (orange) and 
NLL results (black) setting 
$\nu_{us}=C_A\alpha(\nu)/r$ and $\nu=4/r$.\label{FigVbSD}  }  
\end{center}
\end{figure}

Next we consider $E_b$. The situation here is far from 
optimal, as illustrated in \fig{FigVbSD}. We have one less lattice data point 
(the one at shortest distance) than for $E^{(1,0)}$, which makes 
the region, where we can compare with the lattice data, even 
smaller. There are also visible finite size effects in the lattice data, 
when comparing the points for different values of $\beta$. 
This indicates that a global negative shift of our 
lines would be appropriate in the comparison with the lattice data.
We do not implement this shift explicitly in \fig{FigVbSD}, but it is 
understood in the following discussion. 
With fixed $\nu$ the agreement is reasonable. For $\nu \approx 1.5 
\times 1/0.371627$ $r^{-1}_0$ the NLO result agrees well with the data 
and so does the LO result, as there are only small differences 
between the two curves.
Choosing $\nu \sim 1/r$ deteriorates the agreement. 
As for the $1/m$ quasi-static energy, setting $\nu \le 1/r$
causes the perturbation series to blow up (maybe due to renormalons) as the 
scale becomes too small. 
The complete resummation of logarithms yields strongly scale-dependent results. 
If one takes $\nu_{us}=C_A\alpha(\nu)/r$ and 
$\nu=4/r$ the agreement with the data is very good at NLL but 
the difference to the LL prediction is very large. 
Also, we observe a strong dependence of the result on $x_{us}$ (not 
displayed) in the range between one and two.
In any case the uncertainties are large.

\begin{figure}[t]
\begin{center}      
\includegraphics[width=\columnwidth]{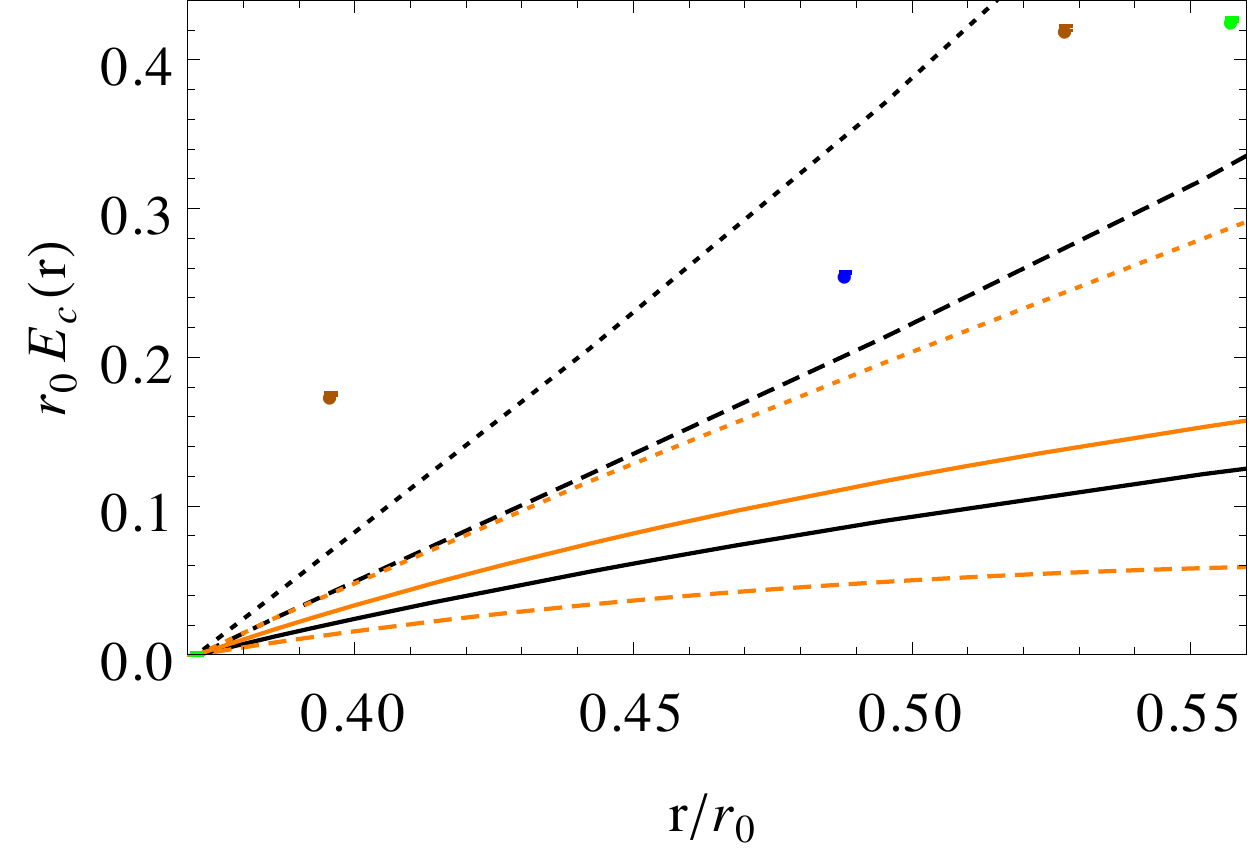}
\caption{Comparison to the lattice data for $E_c$. The solid 
lines correspond to the LO (orange) and NLO results (black) at fixed 
scale $\nu=2\times 1/0.371627 r_0^{-1}$. 
The dashed lines correspond to the LO (orange) and NLO (black) results 
setting $\nu=1/r$. The dotted lines correspond to the LL (orange) and NLL 
(black) results setting $\nu_{us}=C_A\alpha(\nu)/r$ and $\nu=1/r$.} 
\label{FigVcSD}  
\end{center}
\end{figure}

For $E_c$, the situation is similar as for $E_b$ as far as the number and location of the lattice points are concerned, see \fig{FigVcSD}. Note that there are again
sizable finite size effects in the lattice data. 
So, it would be reasonable to add a global shift upwards to all our lines 
when making the comparison with data. 
Working at fixed order with fixed scale the slope is 
smaller than predicted by the data, as can be seen in \fig{FigVcSD} for 
$x=2$. 
Setting $\nu \sim 1/r$ does not help much, until we set 
$\nu \approx 1/r$, where the NLO result gets quite close to the lattice data (unlike 
in the previous cases here we can set $\nu$ quite low). However, it still 
shows a huge difference to the LO result.
Similarly, the resummation of ultrasoft logarithms does not improve 
the situation, unless we choose $x \approx x_{us} \approx 1$, where one 
can get good agreement with the data. 
Nevertheless, once more the scale dependence is large.

\begin{figure}[t]
\begin{center}      
\includegraphics[width=\columnwidth]{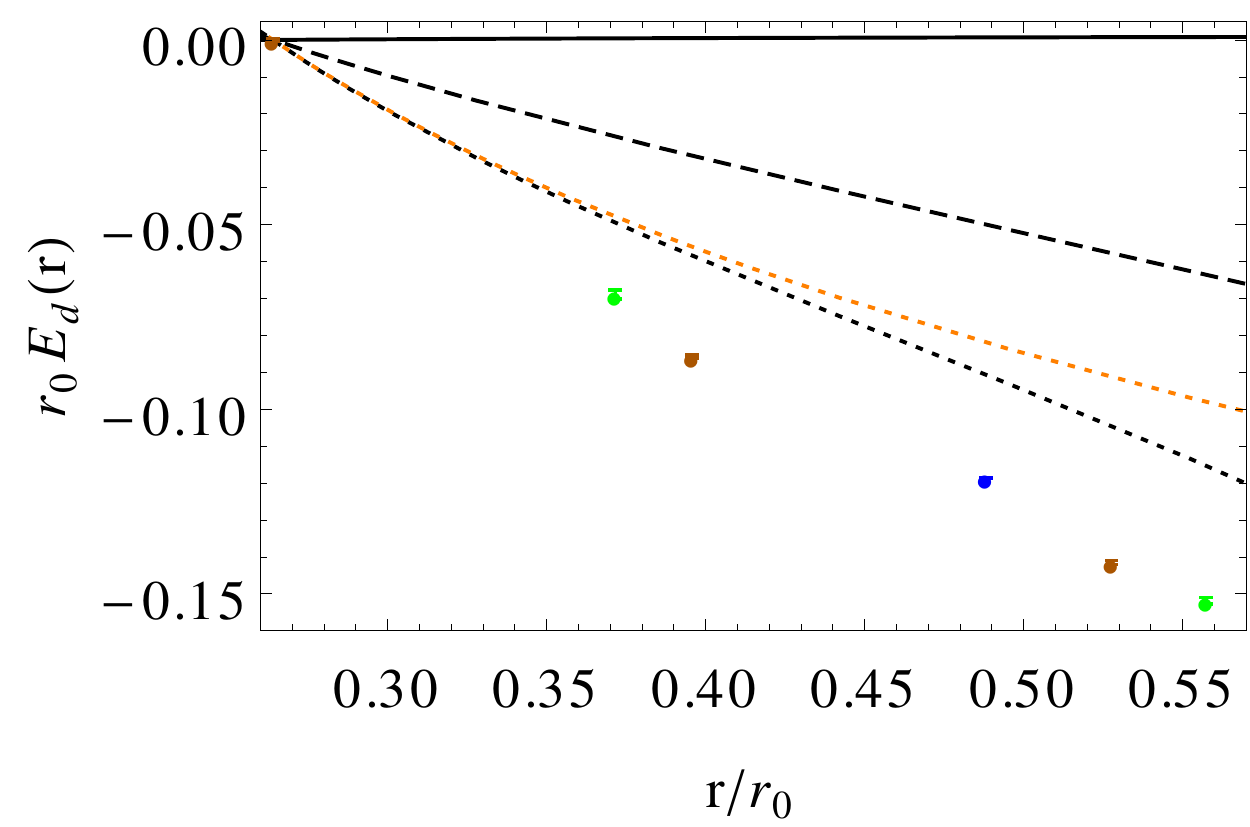}
\caption{
Comparison to the lattice data for $E_d$. The solid 
black line corresponds to the NLO result at fixed scale $\nu=2/0.263821  
r_0^{-1}$. 
The dashed black line corresponds to the NLO result setting $\nu=2/r$. 
The dotted lines correspond to the LL (orange) and NLL (black) results 
setting $\nu_{us}=C_A\alpha(\nu)/r$ and $\nu=2/r$. \label{FigVdSD}
}
\end{center}
\end{figure}

We now turn to $E_d$ and $E_e$. 
They are linear combinations of $E_{\textbf L^2}^{(2,0)}(r)$ and 
$E_{\vp^2}^{(2,0)}(r)$. This makes them particularly interesting, as they vanish 
at $\ord(\al)$. Previous lattice determinations obtained zero slopes 
within uncertainties~\cite{Bali:1997am}. 
Nevertheless, in a dedicated series of works~\cite{Koma:2007jq,Koma:2009ws}, 
the SI momentum-dependent quasi-static energies have been 
computed with increased precision on the lattice and a non-trivial 
behavior has been found at short distances. 
A non-zero slope is also predicted by our computation of the 
quasi-static energies. Therefore,
it is very interesting to check whether they follow the same 
pattern, as we will do in the following. 

For $E_d$ the agreement, if we choose a $r$-independent scale $\nu$, is 
bad. The hardly visible slope of the corresponding curve in \fig{FigVdSD} 
is by far not enough to match the lattice data.
The asymptotic behavior is however correct (in sign) assuming that the 
ultrasoft log dominates, which is not the case for the value of $\nu$ used 
in \fig{FigVdSD} though.
If we set $\nu=2/r$ the agreement is much better. (It gets even 
better for $x \approx 1$). 
The resummation of ultrasoft logs for reasonable values of $x$ and 
$x_{us}$ (optimally $x \approx 2$ while $2 \ge x_{us} \ge 1$) further improves 
the agreement.
However, the observed scale dependence varying $x$ between one and two is 
relatively large.

\begin{figure}[t]
\begin{center}      
\includegraphics[width=\columnwidth]{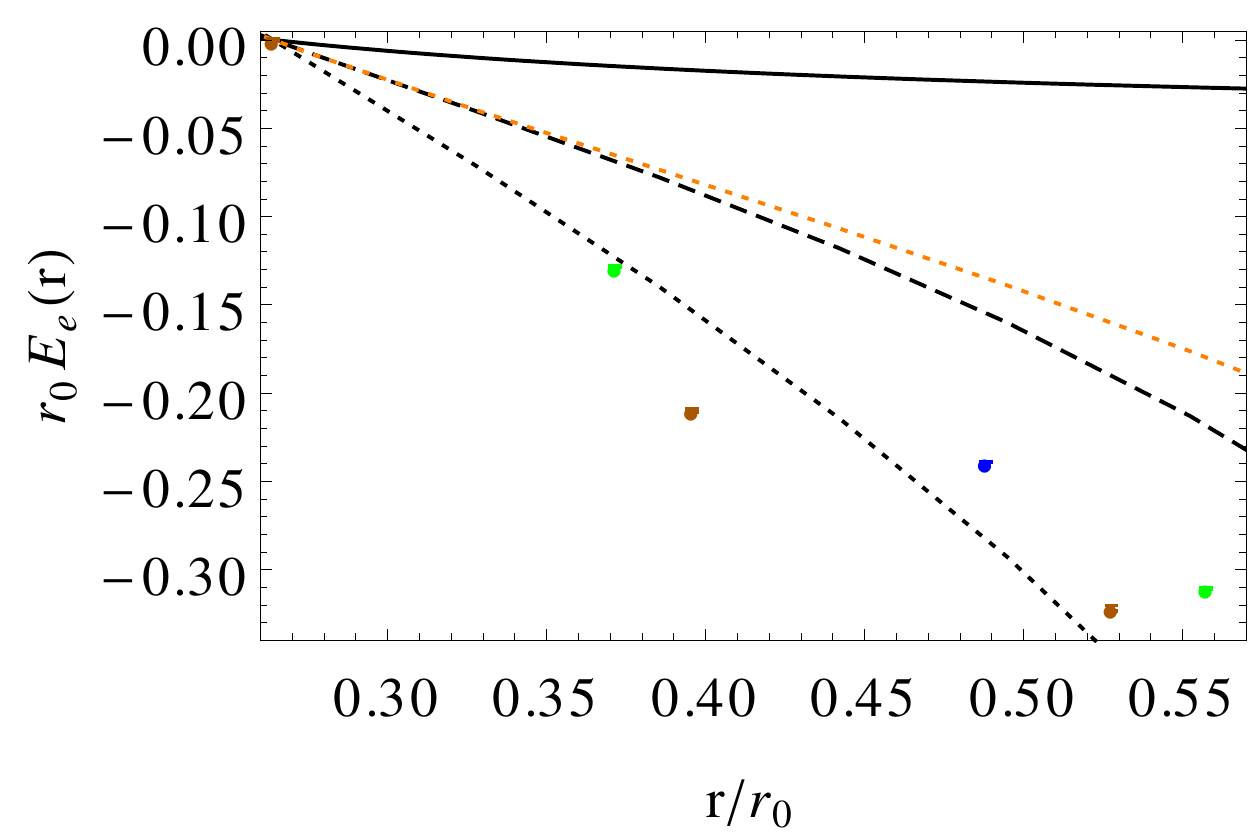}
\caption{Comparison to the lattice data for $E_e$. The solid 
black line corresponds to the NLO result at fixed scale $\nu=2/0.263821 
r_0^{-1}$. 
The dashed black line corresponds to the NLO result setting $\nu=1/r$. 
The dotted lines correspond to the LL (orange) and NLL (black) 
results setting $\nu_{us}=C_A\alpha(\nu)/r$ and $\nu=1/r$.
\label{FigVeSD}} 
\end{center}
\end{figure}

The situation for $E_e$ is similar. Setting $\nu$ constant the agreement is 
bad, as can be seen in \fig{FigVeSD}. If  we set $\nu=1/r$ the 
agreement is much better (optimal for $x\approx 0.83$). One can also 
find good agreement if we resum the ultrasoft logs for reasonable 
values of $x$ and $x_{us}$. Again, the uncertainties due to scale variations in 
the $r$ range where lattice data is available are sizable and prevent a more 
quantitative analysis.

\medskip

Overall, we find that for all studied observables our predictions can be made compatible with the available lattice data. With this statement we mean that for (more or less) reasonable values of the factorization scale we can get reasonable agreement with the lattice data, and/or that the expected size of higher order corrections is large enough to accommodate the difference to the lattice data. We cannot make quantitative 
statements though. Whereas adjusting the factorization scale or incorporating the resummation of logarithms can improve the agreement with the lattice data in some cases, we do not observe a general pattern for all potentials.
Estimates of higher order effects from scale variations 
produce large effects, related to the fact that we have evaluated our 
predictions at quite low scales (large distances). Also, as already 
mentioned, the lattice data is rather scarce. Still, we find it quite 
remarkable that the asymptotic behavior (the sign of the slope) predicted 
by the ultrasoft logarithms $\propto \ln(C_A \al)$ in $E_d$ and $E_e$ complies 
with the  lattice data.

\subsection{Long distances}

\begin{figure}[t]
\begin{center}      
\includegraphics[width=\columnwidth]{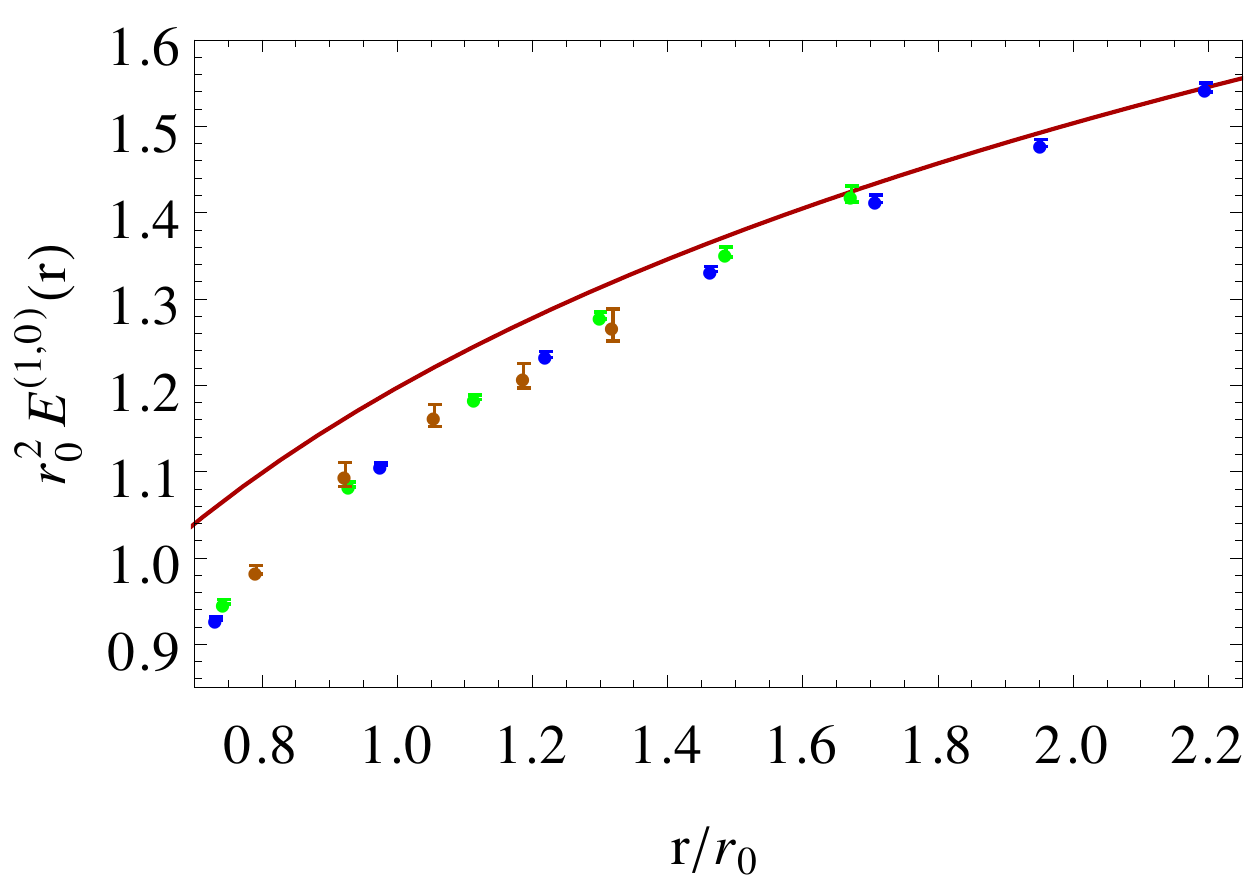}
\caption{Comparison of the lattice data with the predictions from 
effective string theory at long distances for $E_1^{(1,0)}$. \label{FigV10LD}}  
\end{center}
\end{figure}

\begin{figure}[t]
\begin{center}      
\includegraphics[width=\columnwidth]{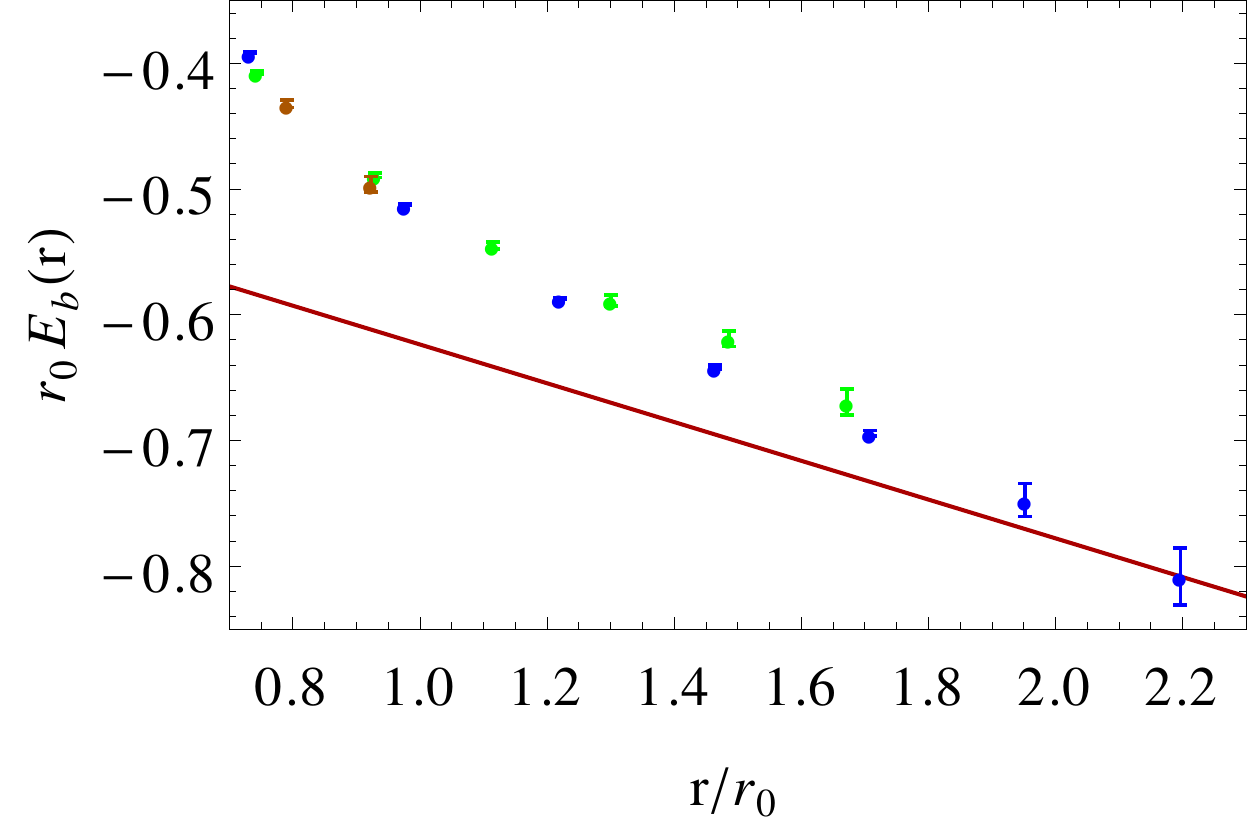}
\includegraphics[width=\columnwidth]{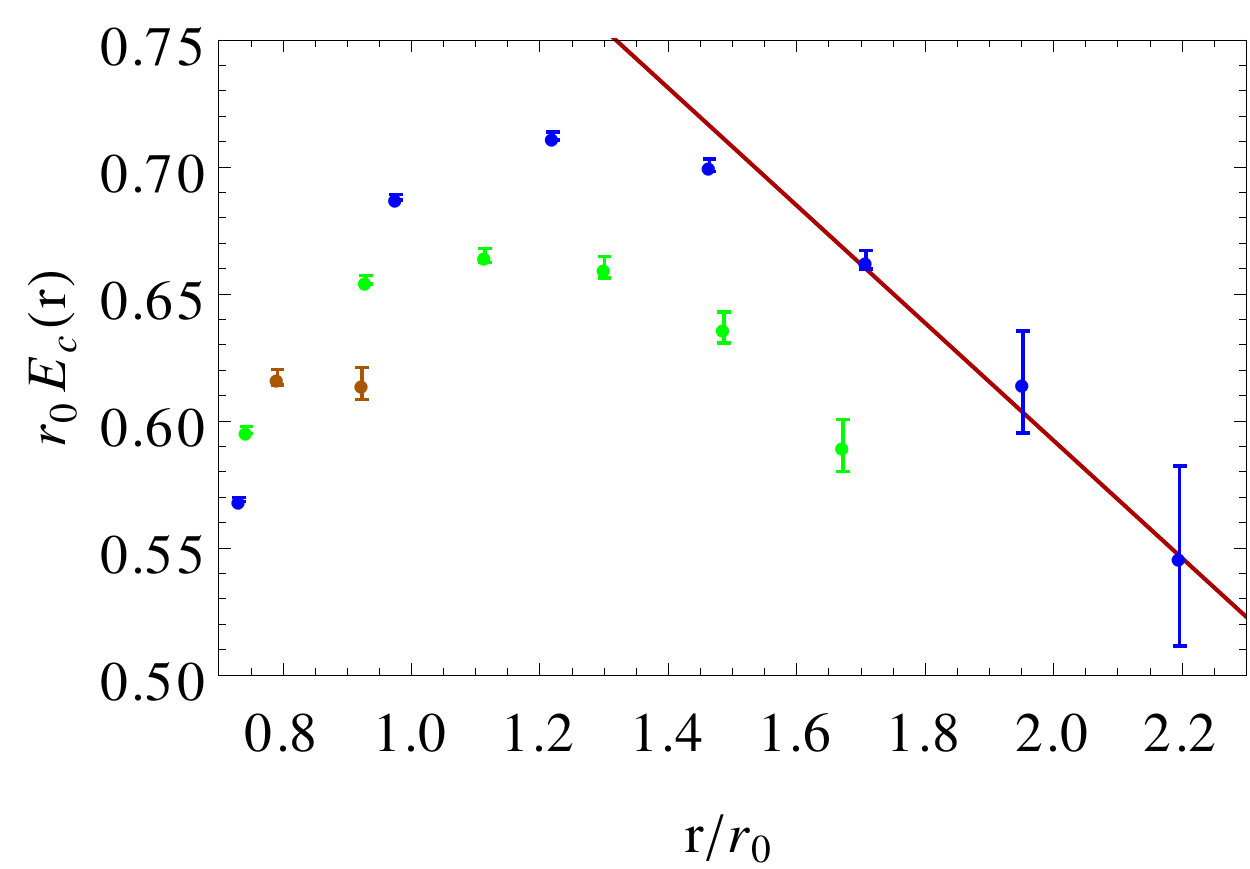}
\caption{Comparison of the lattice data with the predictions from 
effective string theory at long distances for $E_b$ and 
$E_c$. \label{FigVbcLD}}  
\end{center}
\end{figure}

It is also interesting to consider the long-distance behavior of the 
quasi-static energies.  Their behavior cannot be predicted by QCD 
using analytic tools. Still, it can be simulated by computations in 
the discretized version of QCD, i.e. on the lattice. 
One can then compare these numerical results with model 
predictions. Here we focus on the effective string theory predictions for the 
$1/m$ and the SI momentum-dependent quasi-static energies 
obtained in  \rcite{PerezNadal:2008vm}. For the SI 
momentum-dependent quasi-static energies, they are actually equivalent to the 
predictions obtained first with the minimal area law~\cite{Barchielli:1986zs}.%
\footnote{Other models also share the same behavior in the long-distance limit, 
see the discussion in \rcite{Brambilla:1996aq}.} 
The latter however misses the $1/m$ potential. 
The string model gives the following results at leading 
order in the $1/(r\lQ)$ expansion:
\begin{align}
E^{(1,0)}&=\frac{\sigma}{2\pi}\ln(\sigma r^2)+\mu_1,\\
E_b(r)&= E_e(r)=-\frac{r \sigma}{9},\\
E_c(r)&=E_d(r)= -\frac{r \sigma}{6}.
\end{align}
These expressions only depend on the string tension $\sigma$.%
\footnote{The dependence on $\mu_1$ that appears in $E^{(1,0)}$ vanishes 
when comparing with lattice simulations as we will only compare differences of static energies 
evaluated at different $r$.} 
The string tension can be obtained from fits to the static potential at long 
distances. The number we use is taken from 
\cite{Necco:2001xg} and reads: $\sigma=1.3882r_0^2$. Therefore, the above 
expressions are in fact predictions. On the lattice side, to obtain 
absolute normalizations is problematic. Hence, just like at short 
distances we compare energy differences. We choose to 
normalize our prediction to the longest distance (rightmost) lattice 
point available, where the above models are expected to work best. 
The comparison with the lattice is shown in \figss{FigV10LD}{FigVbcLD}{FigVdLD}. There is  
qualitative agreement between the lattice and the effective string theory 
predictions.%
\footnote{In \fig{FigVbcLD} it is obvious 
that there are still sizable finite size effects as there is no good 
scaling between the lattice data points for $\beta=5.85$ (blue) and 
$\beta=6$ (green). } 
On the other hand it is evident that the agreement is far from perfect and the 
slopes demand for corrections to the leading string behavior. Such corrections 
can in principle be computed in the same way as for the static 
potential, where they give rise to the L\"uscher term~\cite{Luscher:1980fr}. 
It would be interesting to see if the analogous computations for the 
quasi-static energies (in case they happen to be free of new constants) 
improve the agreement with the lattice data. 

\begin{figure}[t]
\begin{center}      
\includegraphics[width=\columnwidth]{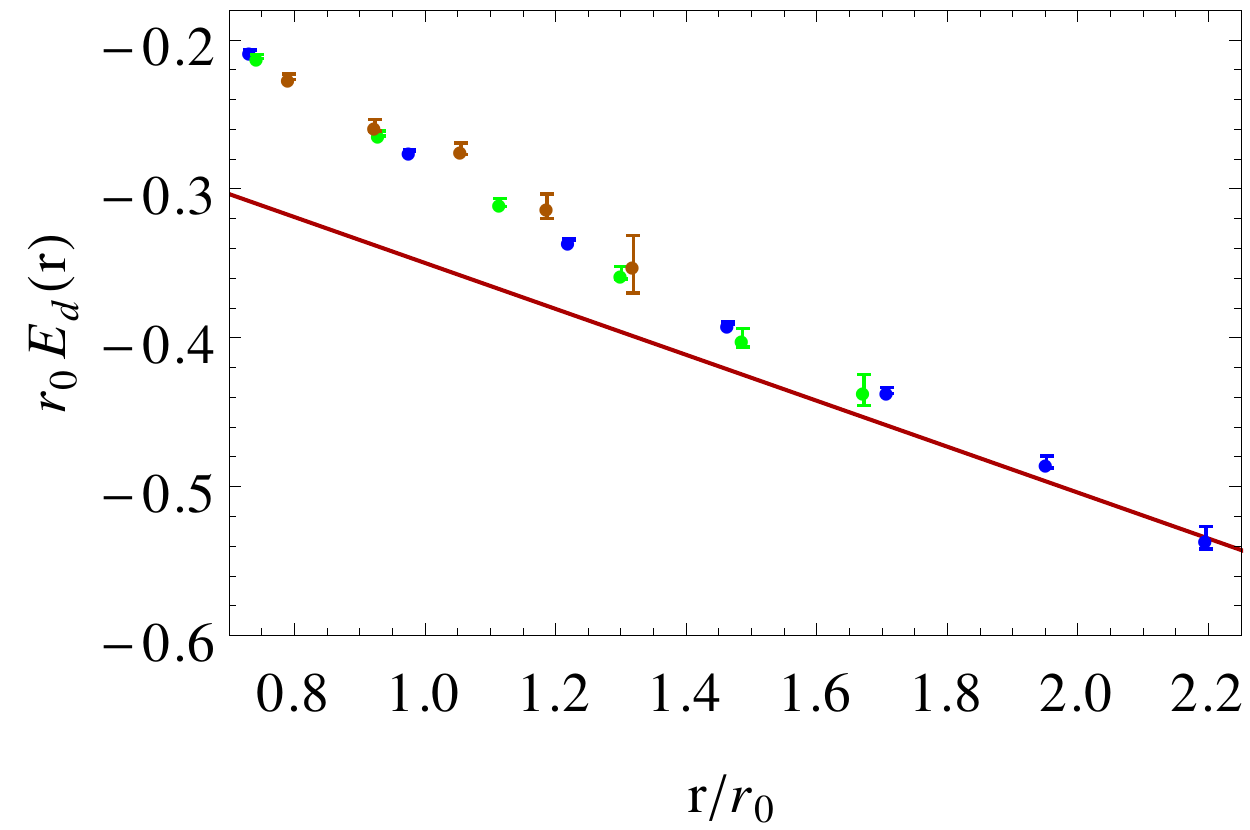}
\medskip
\includegraphics[width=\columnwidth]{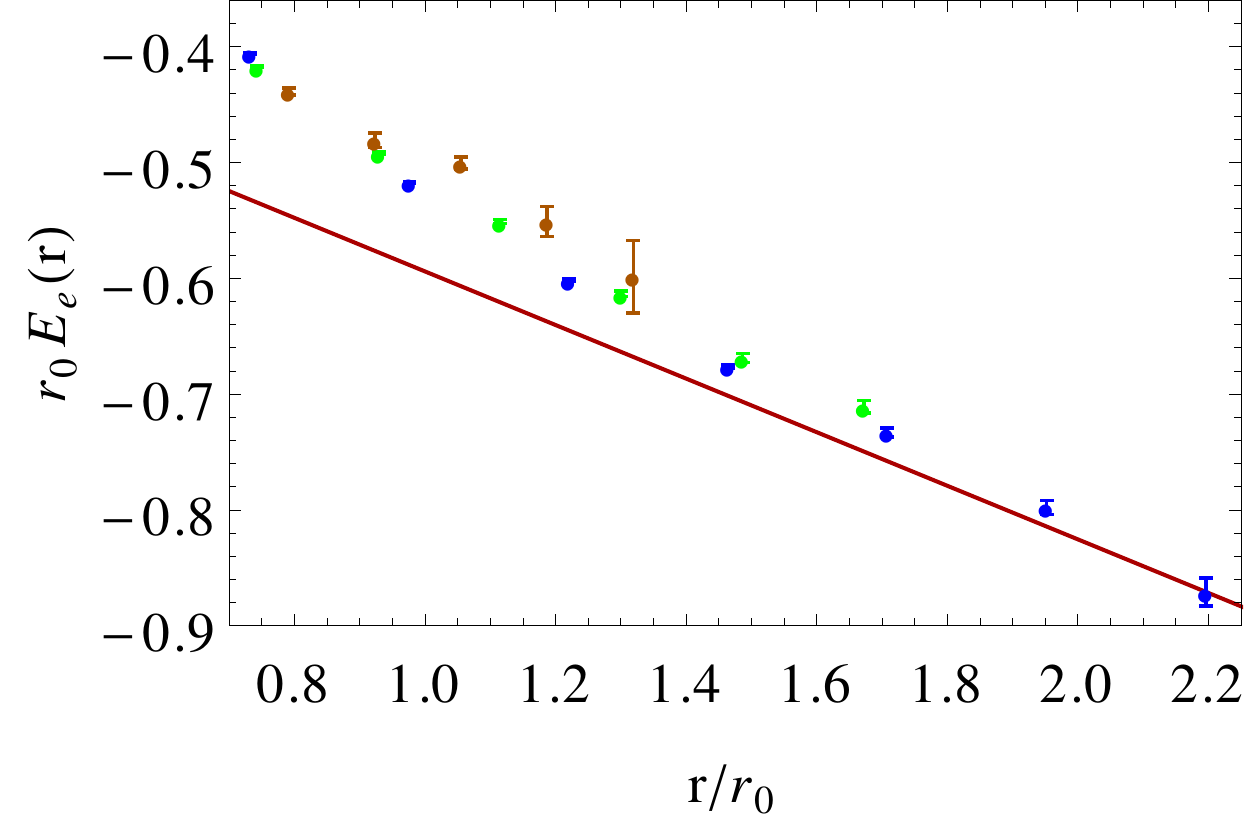}
\caption{Comparison of the lattice data with the predictions from 
effective string theory at long distances for $E_d$ and $E_e$. \label{FigVdLD}}
\end{center}
\end{figure}
\section{Conclusions}

In this paper we have determined the short-distance behavior of the 
nonperturbative generalizations of the $1/m$ potential and of the $1/m^2$ 
SI momentum-dependent heavy quarkonium potentials, 
in terms of Wilson loops (as defined in 
\Sec{sec:E}), with $\ord(\al^3)$ and $\ord(\al^2)$ accuracy, respectively. 
We refer to these objects as quasi-static energies. 
Our results can be found in \eqss{E10SD}{EL11SD}. 
For our (NLO) computation it is crucial that the quasi-static energies 
start to be sensitive to the ultrasoft scale $\sim C_A \al/r$.
Therefore, their calculation requires a resummation of an infinite number 
of potential interactions to account for the ultrasoft effects.

We have computed these ultrasoft contributions to the quasi-static 
energies in a diagrammatic approach. This included Feynman graphs with one 
ultrasoft and up to three potential loops. We expect that a proper EFT setup will significantly simplify such computations.
At the same time this would allow for a systematic RG analysis beyond LL 
precision.
With our method we were able perform the correct LL resummation 
of ultrasoft logarithms, but can only provide an educated guess for the NLL 
results.

Our results should reproduce the short-distance limit of lattice 
simulations of the same quasi-static energies. 
Some of these were evaluated long ago in \rcite{Bali:1997am}, and more 
recently in \rcites{Koma:2006si,Koma:2007jq,Koma:2009ws}. 
We have performed a preliminary comparison with the lattice 
data~\cite{Koma:2006si,Koma:2007jq,Koma:2009ws} at short distances in 
\Sec{Sec:lattshort}. As there is not enough lattice data at short distances, we 
were unable to perform a quantitative analysis. We had to compare at 
rather low scales. This causes large uncertainties from scale 
variations. Within these uncertainties we can find qualitative agreement with 
the data. 
It is quite remarkable that the sign of the slopes 
obtained by the lattice simulations agrees with the asymptotic behavior 
predicted by our weak coupling analysis. 
Whereas for three of the considered quasi-static energies, the 
behavior is dominated by the soft LO contribution, for two of 
them the sign of the slope is dictated by the ultrasoft contribution. 
This makes them particularly interesting.  
Those quasi-static energies vanish at 
$\ord(\al)$, but are non-zero at $\ord(\al^2)$. 
Earlier lattice simulations predicted an approximately zero slope for them in the short distance limit. 
The most recent simulations show a non-zero behavior consistent with our 
results, albeit within large uncertainties. 

We also compare the lattice data with the expectations from effective string 
models at long distances. We find qualitative agreement, but 
there is obvious room for improvement. 
In the long term one could try to study (qualitatively) how the short and 
long distance predictions (for a specific model) combine.

\medskip

{\bf Acknowledgments} \\
We thank Yoshiaki and Miho Koma for providing us with the lattice data of 
\rcites{Koma:2007jq,Koma:2009ws}. A.P. thanks Gunnar S. Bali for discussions.
M.S. is supported by GFK and by the PRISMA cluster of excellence at JGU Mainz.
This work was supported in part by the Spanish grants FPA2014-55613-P, 
FPA2013-43425-P and SEV-2012-0234, and the Catalan grant
SGR2014-1450.

\appendix

\section{Results for the ultrasoft diagrams}
\label{FGresults}

For each diagram $X$ in \fig{Vx0Diags} (and equivalently in 
\fig{V11Diags}) we have computed the object ${\cal W}_n^{X,ij}$ according 
to \eq{Wdef}, in both Coulomb and Feynman gauge. 
The bare results  in  Feynman gauge (FG) read:
\begin{align}
{\cal W}_{n,\rm FG}^{1a,ij}&=
(-i)^{n+1} \frac{C_F}{2} \, g^2  \delta^{ij} \Delta V^{D-n-1} 
(D-1) \pi ^{-D/2} \nn\\
 &\times\Gamma \left(\frac{D}{2}\right)  \Gamma (1-D+n)
\,,\\
{\cal W}_{n,\rm FG}^{1b,ij}&=
-(-i)^{n+1} \frac{C_F}{2} \, g^2 \delta^{ij} \Delta V^{D-n-1}
 \pi ^{-D/2} \nn\\
 &\times\Gamma \left(\frac{D}{2}\right)  \Gamma (1-D+n)
\,,
\\
{\cal W}_{n,\rm FG}^{2a,ij}&=
(-i)^{n+1} C_A C_F \, g^4 \Big((D-3) {\hat {\bf r}}^i{\hat {\bf r}}^j-\delta^{ij}\Big) 
r^{3-D}\nn\\
 &\hspace*{-0.5cm} \times\Delta V^{D-n-2}
(D-2) 2^{-D} \pi ^{1-D} \Gamma (D-3) \Gamma (2-D+n) 
\,,\\
{\cal W}_{n,\rm FG}^{2b,ij}&=
(-i)^{n+1} C_A C_F \, g^4  \delta^{ij}  r^{3-D} \Delta V^{D-n-2} 
(D-2) \nn\\
 &\times
2^{-D}  \pi ^{1-D} \Gamma (D-3) \Gamma (2-D+n)
\,,\\
{\cal W}_{n,\rm FG}^{2c,ij}&=
-(-i)^{n+1} C_A C_F \, g^4 {\hat {\bf r}}^i{\hat {\bf r}}^jr^{3-D} \Delta V^{D-n-2}\nn\\
 &\times
\frac{2^{-D} \pi ^{1-D}}{n+1}  \Gamma (D-1) \Gamma (2-D+n)
\,,\\
{\cal W}_{n,\rm FG}^{3a,ij}&=
-(-i)^{n+1} C_A^2 C_F \, g^6 
\Big((D-5) (D-3) {\hat {\bf r}}^i{\hat {\bf r}}^j +\delta^{ij}\Big)\nn\\
 &\times r^{6-2 D} 
\Delta V^{D-n-3}
4^{-D} \pi ^{1-\frac{3 D}{2}}\sin \left(\frac{\pi  D}{2}\right) \nn\\
&\times 
\Gamma \left(2-\frac{D}{2}\right) \Gamma^2 (D-3)  
\Gamma (3-D+n) 
\,,\\
{\cal W}_{n,\rm FG}^{3b,ij}&=
-(-i)^{n+1} C_A^2 C_F \, g^6 \delta^{ij} r^{6-2 D} \Delta V^{D-n-3}
4^{-D} \pi ^{1-\frac{3 D}{2}} \nn\\
&\times \sin \left(\frac{\pi D}{2}\right) 
\Gamma \left(2-\frac{D}{2}\right) 
\Gamma^2 (D-3) \Gamma (3-D+n)
\,,\\
{\cal W}_{n,\rm FG}^{3c,ij}&=
-(-i)^{n+1} C_A^2 C_F \, g^6 \Big((D-3) {\hat {\bf r}}^i{\hat {\bf r}}^j -\delta^{ij} \Big)  
r^{6-2 D} \nn\\
&\times\Delta V^{D-n-3}
2^{1-2 D} \pi ^{1-\frac{3 D}{2}} \nn\\
&\times
\sin \left(\frac{\pi  D}{2}\right) 
\Gamma \left(2-\frac{D}{2}\right) \Gamma^2 (D-3)  
\Gamma (3-D+n) 
\,,\\
{\cal W}_{n,\rm FG}^{3d,ij}&=
-(-i)^{n+1} C_A^2 C_F \, g^6 {\hat {\bf r}}^i{\hat {\bf r}}^jr^{6-2 D} \Delta V^{D-n-3}
\nn\\
&\times \frac{4^{1-D} \pi ^{1-\frac{3 D}{2}}}{n+1} 
\sin \left(\frac{\pi  D}{2}\right) \Gamma \left(3-\frac{D}{2}\right)\nn\\
&\times \Gamma 
(D-4) 
\Gamma (D-1)  \Gamma (3-D+n)
\,,\\
{\cal W}_{n,\rm FG}^{3e,ij}&=
(-i)^{n+1} C_A^2 C_F \, g^6 {\hat {\bf r}}^i{\hat {\bf r}}^j r^{4-2 D} \Delta V^{D-n-5}\nn\\
&\times
\frac{4^{-D} \pi ^{1-\frac{3 D}{2}}
}{(n+1) (n+2)} \sin \left(\frac{\pi  D}{2}\right) \Gamma 
\left(2-\frac{D}{2}\right) \nn\\
& \times
\Gamma^2 (D-2) \Gamma (3-D+n) 
\nn\\
&\times\big[(D-2) \Delta V^2 r^2-2 (D-n-4)(D-n-3)\big]
\,,\\
{\cal W}_{n,\rm FG}^{3f,ij}&=
(-i)^{n+1} C_A^2 C_F \, g^6 {\hat {\bf r}}^i{\hat {\bf r}}^j r^{4-2 D} \Delta V^{D-n-5}
\frac{4^{-D} \pi ^{1-\frac{3 D}{2}}}{\Gamma (n+3)} \nn\\
&\times
\sin \left(\frac{\pi  D}{2}\right) \Gamma \left(2-\frac{D}{2}\right) \Gamma^2 
(D-2) \nn\\
&\times
\Gamma (n+1)  \Gamma (5-D+n)
\,,\\
{\cal W}_{n,\rm FG}^{3g,ij}&={\cal W}_{n,\rm FG}^{3f,ij}
\,,\\
{\cal W}_{n,\rm FG}^{4a,ij}&=
-(-i)^{n+1} C_A^3 C_F \, g^8 {\hat {\bf r}}^i{\hat {\bf r}}^j r^{9-3 D} \Delta V^{D-n-4}
\nn\\
&\times \frac{2^{2-3 D} \pi ^{1-2 D}}{n+1} 
\sin^2 \left(\frac{\pi D}{2}\right) \nn\\
&\times
\Gamma^2  \left(2-\frac{D}{2}\right)\Gamma^2 (D-3) \Gamma (D-2)  \Gamma(4-D+n)
\,,\\
{\cal W}_{n,\rm FG}^{4b,ij}&=
-(-i)^{n+1} C_A^3 C_F \, g^8 {\hat {\bf r}}^i{\hat {\bf r}}^j r^{9-3 D} \Delta V^{D-n-4}
\nn\\
& \times
\frac{2^{4-3 D} \pi ^{1-2 D} }{n+1} \sin^2 \left(\frac{\pi D}{2}\right) \Gamma^2 \left(3-\frac{D}{2}\right)\nn\\
&\times 
\Gamma (D-4) \Gamma (D-3) \Gamma (D-2) \Gamma (4-D+n)
\,,\\
{\cal W}_{n,\rm FG}^{5a,ij}&=
-(-i)^{n+1} C_A^4 C_F \, g^{10} {\hat {\bf r}}^i{\hat {\bf r}}^j r^{12-4 D} \Delta V^{D-n-5}
\nn\\
&\times \frac{2^{4-4 D} \pi ^{1-\frac{5 D}{2}} \Gamma (n+1)}{\Gamma (n+3)} 
\sin^3\left(\frac{\pi  D}{2}\right)
\Gamma \left(2-\frac{D}{2}\right) \nn\\
&\times
\Gamma^2 \left(3-\frac{D}{2}\right) \Gamma^2 (D-4) 
\Gamma^2 (D-2)  \Gamma (5-D+n)
\,,
\end{align}
where $D=d+1=4+2\epsilon$ and ${\hat {\bf r}} = {\bf r}/r$. Using \eq{DeltaVLO} 
the sum of the ${\cal 
W}_{n,\rm FG}^{X,ij}$ gives the gauge invariant result in \eq{TotWn}.

In Coulomb gauge (CG) we obtain:
\begin{align}
{\cal W}_{n,\rm CG}^{X,ij} &= \frac{D-2}{D-1} \;{\cal W}_{n,\rm FG}^{X,ij} \nn\\
&\text{for} \quad 
X=\{1a, 2a, 2b, 3a, 3b, 3c, 3d, 4a, 4b, 5a\}\,,\\
{\cal W}_{n,\rm CG}^{Y,ij} &= 0 \nn\\
 &\text{for} \quad 
Y=\{1b, 2c, 3e, 3f, 3g\}\,.
\end{align}
The fact that the sum of the ${\cal W}_{n,\rm CG}^{X,ij}$ gives again the gauge invariant result in \eq{TotWn} is a strong cross check of our computation.

\end{document}